\documentclass[12pt]{JHEP3}
\pdfoutput=1
\usepackage{amsmath,amssymb}
\usepackage{yfonts}
\usepackage[pdftex]{graphicx}

\newcommand{\be}{\begin{equation}}
\newcommand{\ee}{\end{equation}}
\newcommand{\beqa}{\begin{eqnarray}}
\newcommand{\eeqa}{\end{eqnarray}}
\newcommand{\dA}{{\cal{A}}}

\newcommand{\dZ}{{\cal Z}}

\def\bemat{\left( \begin{array}}
\def\enmat{\end{array} \right)}

\newcommand{\qn}{\textswab{q}}
\newcommand{\wn}{\textswab{w}}

\newcommand{\im}{\hbox{\,Im\,}}
\newcommand{\kk}{\bf k}

\newcommand{\tpsi}{\tilde \psi }
\newcommand{\Dis}{{D}}
\newcommand{\tr}{{\rm tr}}

\def\med{\frac{1}{2}}
\def\d{\partial}

\relax

\title{Holographic Spectral Functions at Finite Baryon Density}

\author{ Javier Mas\footnote{jamas@fpaxp1.usc.es}, Jonathan Shock\footnote{shock@fpaxp1.usc.es}, Javier Tarr\'\i o\footnote{tarrio@fpaxp1.usc.es} and Dimitrios Zoakos\footnote{zoakos@fpaxp1.usc.es}
\\
 Departamento de F\'\i sica de Part\'\i culas,
Universidade de Santiago de Compostela \\ 
and\\
Instituto Galego de F\'\i sica de Altas Enerx\'\i as (IGFAE)\\
E-15782 Santiago
de Compostela, Spain\\
}

\abstract{
Using the AdS/CFT correspondence, we compute the spectral functions of thermal  super Yang Mills  at  large $N_c$ coupled to a small number of flavours of fundamental matter, $N_f\ll N_c$, in the presence of a nonzero baryon density. The holographic dual of such a theory involves the addition of probe D7-branes with a background worldvolume gauge field switched on,  embedded in the geometry of a stack of black D3-branes. We perform the analysis in the vector and scalar channels which become coupled for nonzero values of the spatial momentum and baryon density. In addition, we obtain the effect of the 
presence of net baryon charge on the photon production. We also extract the conductivity and find perfect agreement with the results derived by Karch and O'Bannon in a macroscopic setup.
}

\begin{document}


\section{Introduction}

Holographic techniques based on the AdS/CFT correspondence  \cite{hep-th/9711200,hep-th/9905111} have become a fruitful arena in which to address questions concerning properties of a strongly coupled non-abelian plasma (see \cite{arXiv:0804.2423} and references therein). 
The finite temperature scenario \cite{hep-th/9803131} shares many more properties with what is expected from 
thermal QCD than its zero temperature counterpart. Within this context, 
holographic spectral functions \cite{hep-th/0602059} have been the subject of much attention in the last year \cite{arXiv:0706.0162,arXiv:0709.2168,arXiv:0710.0334, Amado:2007yr}.  They carry information about transport coefficients such as the conductivity and viscosity of the plasma  \cite{hep-th/0405231} as well as about particle production rates  \cite{McLerran:1984ay}. This is very appealing, since photon and dilepton production are amongst the most interesting signatures  of a quark-gluon plasma.
In fact these two phenomena have been studied in a number of
recent papers starting with \cite{hep-th/0607237} where the geometry is that of a stack of black D3-branes, corresponding to pure ${\cal N}=4$ SYM at finite temperature with a weakly gauged  $U(1)_R$ subgroup
of $SU(4)_R$, under which the  ${\cal N}=4$ fermions and scalar fields
are charged. It was shown that in order to calculate the
photon and dilepton emission rates, at small electromagnetic coupling,
the dependence on the $U(1)$ gauge field is subleading in the expression for
current-current correlators and, therefore, the calculation can be
performed solely in reference to the non-abelian gauge fields, with no need for
a dynamical photon. This simplifies the problem greatly as the
full gravity dual of ${\cal N}=4$ SYM coupled to weakly gauged electromagnetism is
not known.

A step towards a more realistic model of QCD required the presence of
charged  fundamental matter. This technology is now well under control
in the quenched approximation, and involves the introduction of $N_f \ll N_c$ probe
branes in the dual gravitational background \cite{hep-th/0205236}.  
The study of this framework has revealed an interesting phase structure, in which
the adjoint and fundamental matter deconfine at different temperatures
 \cite{hep-th/0306018,hep-th/0611021,hep-th/0611099}.
Photoemission in such setups  has been
recently studied in \cite{arXiv:0709.2168} in the D3/D7 and the D4/D6
systems, in \cite{hep-th/0610247} in the context of the
Sakai-Sugimoto model \cite{hep-th/0412141}, and in \cite{arXiv:0802.1460} for the AdS/QCD background.

In order to unravel the phase diagram  of QCD-like theories it is 
mandatory to go beyond the vanishing chemical potential limit, $\mu=0$ \cite{hep-ph/0509068,hep-ph/0011333}.
 Recently, the study of the strongly coupled deconfined plasma from AdS/CFT with
nonzero baryon or isospin chemical potential has become more than an academic exercise \cite{arXiv:0710.0334,arXiv:0709.3948,hep-th/0611099,arXiv:0705.3870,arXiv:0708.2818,arXiv:0708.3706,arXiv:0709.0570,arXiv:0709.1225,arXiv:0711.0407,arXiv:0709.3948}. 
For instance, at low temperatures and high densities one expects to find 
interesting new phases, like the color-flavor locked  superconductor, that 
could occur in the interior of very dense neutron stars.  At large $N_c$ a number of new phases have been proposed in \cite{arXiv:0706.2191,arXiv:0803.0279} which may have implications for  real  QCD.
Concerning heavy ion experiments, the phenomenological  appeal of this extension is less clear. While in the fireball the residual baryon density at the core is tiny, in experiments at SPS, fits are nonetheless consistent
with values of the chemical potential of $\mu_0\sim 400 MeV$ \cite{nucl-th/0511071,hep-ph/0511094}. Whether this  will be enough to locate the
critical point between a crossover and first order phase transition is still unclear.
Therefore one ought to address issues like photon and dilepton emission as a means of gaining
insight into different regions of the phase diagram.

The present work is an extension of \cite{arXiv:0709.2168} to include both chemical potential and finite spatial momentum. Note also that in \cite{arXiv:0710.0334,arXiv:0802.1460,arXiv:0804.2168} the authors made some progress in this direction, although for simplicity they  studied only a subsector of the possible spectral functions.

Next we review the strategy of the computation, as discussed in \cite{hep-th/0607237}.
The basic object to compute is the spectral function, related to the retarded two-point function $G^{R}_{\mu\nu}$  as follows
\be\label{spectdef}
\chi_{\mu\nu}(k) = -2 \im  G^{R}_{\mu\nu}\, .
\ee
At zero temperature, the form of the retarded correlator  is dictated by Lorentz invariance and  gauge symmetry
\be
G^{R}_{\mu\nu} = P_{\mu\nu}(k) \Pi(k^2)\, ,
\ee 
where $P_{\mu\nu} = \eta_{\mu\nu}-k_\mu k_\nu/ k^2$ is the  transverse projector in Minkowski space
and $\eta_{\mu\nu} = \hbox{diag}(-1,1,1,1)$.
In ${\cal N}=4$ supersymmetric Yang Mills  theory, the form of $\Pi(k)$ is dictated by scale invariance up to a constant that can be computed and  yields, at large $N_c$,  \cite{hep-th/0607237}
\be
\chi_{\mu\nu}^{T=0}(k) = P_{\mu\nu}(k) \frac{N_c N_f}{16\pi}|k^2| \Theta(-k^2) \hbox{sgn}(k^0)\, .
\label{zerotemp}
\ee
At nonzero temperature $T\neq 0$ only rotational invariance remains unbroken, and the tensor decomposition of the retarded correlator now defines two polarization tensors $\Pi^\perp(k) $ and $  \Pi^{||}(k)$ as follows
\be
G^R_{\mu\nu}(k) = P^\perp_{\mu\nu}(k)\, \Pi^\perp(k) + P^{||}_{\mu\nu}(k)\, \Pi^{||}(k) \, ,
\ee
with the transverse and longitudinal projectors defined as follows: 
$ P^\perp_{00}(k)=0, P^\perp_{0i}(k) = 0, P^\perp_{ij}(k) = \delta_{ij} - k_ik_j/\kk^2$, $P^{||}_{\mu\nu}(k) + P^\perp_{\mu\nu}(k) =  P_{\mu\nu}(k) =\eta_{\mu\nu}- k_\mu k_\nu/k^2$.
For example if $k^\mu = (\omega,q,0,0)$ we have
\be
P^{\perp}_{22} = P^{\perp}_{33}  = 1~;~~~~
P^{||}_{00}= \frac{q^2}{\omega^2-q^2}~;~~~~
P^{||}_{01}= \frac{-\omega q}{\omega^2-q^2}~;~~~~
P^{||}_{11}= \frac{\omega^2}{\omega^2-q^2}~.~
\ee
Observables, like the conductivity and particle production rates, are encoded in
the  trace of the spectral function, which can decomposed as
\be
\chi^\mu{_\mu}(k) = -4 \im \Pi^\perp(k) - 2 \im \Pi^{||}(k)  \label{trspden} \equiv
\chi^\perp(k) + \frac{1}{2} \chi^{||}(k)
\, .
\ee

For lightlike momenta,
$k^2=0$, we see that $P^{||}_{\mu\nu}$ diverges. Hence, on the light cone, we have that
\be
\lim_{k^0\to |{\bf k}|} \Pi^{||}(k) = 0\, .
\ee
Otherwise we would have a divergence of $G^R$.
Therefore, $\chi^\mu{_\mu}(k) $ at lightlike momentum, is controlled by the transverse polarization $\Pi^\perp$ which is enough to compute  the emission rate for real photons \cite{McLerran:1984ay}
\be
d\Gamma_\gamma = 
\left.\frac{dk^3}{(2\pi)^3}\frac{e^2}{2|{\bf k}|} n_B(k) \chi^\mu{_\mu}(k)\right\vert_{k^0=|{\bf k}|}\, .
\ee
Here, the Boltzmann factor, $n_B(k) = (e^{k_0/T}-1)^{-1}$,  receives no contribution from the
baryon chemical potential because it refers to the bath of thermal photons which have zero baryon number.
When dealing with timelike momenta, 
rotational invariance relates these two polarizations for vanishing three-momentum ${\bf q}$ where
$k^\mu=(\omega,{\bf q})$,
\be
\lim_{{\bf q}\to 0}\Pi^\perp(\omega,{\bf q}) = \lim_{{\bf q}\to 0}\Pi^{||}(\omega,{\bf q}) \, .
\ee
For nonzero ${\bf q}$ both polarizations contribute to  $\chi^\mu{_\mu}(k) $ and encode the
production of virtual photons  which eventually decay into dilepton pairs, $l\bar l$, of momentum
$k^\mu_l + k^\mu_{\bar l} = k^\mu$. If the lepton $l$ has mass $m_l$ and charge $e_l$,
the differential dilepton emission rate per unit four-volume is given by
\be
d\Gamma_{l\bar  l} = \frac{d^4 k}{(2\pi)^4}\frac{e^2 e_l^2}{6\pi |k^2|^{5/2}}
\Theta(k^0)\Theta(- k^2-4 m_l^2)\sqrt{-k^2-4m_l^2}(-k^2+2m_l^2)n_B(k) \chi^\mu{_\mu}(k)  \, .
\ee
As we see, the relevant part of the computation resides in the retarded correlator.  At strong coupling, for the time being, only holographic techniques are easy to implement, albeit in  ${\cal N}=4$ supersymmetric Yang Mills at large $N_c$.
 
 In \cite{hep-th/0612169} the quasinormal spectrum for strongly coupled finite temperature ${\cal N}=4$ SYM was calculated from a holographic perspective. The `melting meson' scenario defines a spectrum of states where the poles in the correlator are found in the complex plane and therefore correspond to modes of finite lifetimes. For this reason, the interpretation of the spectral functions which we will be calculating in the following will be clear in terms of the set of quasinormal mesons in the plasma.

The outline of the paper is as follows. In section \ref{setup} we outline the brane construction of flavor probes in the presence of a finite baryon number density and discuss the regions of interest in the parameter space in terms of the effective horizon area on the probe. In section \ref{fluctuations} we discuss the calculation of the transverse gauge fluctuations on the probe brane and the coupled system involving the longitudinal modes and the scalar perturbations. We show how the spectral function can be calculated in all three sectors. In section \ref{graphressec} we give results in various regions of parameter space for the spectral function, the photoproduction rates and the limiting velocity of the mesons, as calculated from the peaks in the spectral functions. We illustrate how the presence of a finite baryon number and finite spatial momentum affect all   these results. In section \ref{conductivitysec} we discuss how the conductivity can be calculated using the microscopic approach and show that we get the same answer as that calculated in the macroscopic regime in \cite{arXiv:0705.3870}. In section \ref{conclusions} we discuss these results and provide new directions for future research. In the appendices we provide more details of the calculations given in the bulk of the paper as well as providing some exact, analytic results in various regions of momentum space.

\section{Holographic setup}\label{setup}

In the framework of the AdS/CFT correspondence, the retarded correlator $G^R(k)$ can be obtained from the perturbations of a $U(1)$ gauge field dual to the electromagnetic current on the boundary.
 The relevant holographic description is provided by an AdS geometry with a non-extremal horizon and embedded probe branes. The baryonic $U(1)$ symmetry is the abelian center of the natural $U(N_f)$ global symmetry present on a stack of $N_f$ coincident D-branes.
For the case of interest here, namely Dp/Dq configurations, the dynamics of this
gauge field is fully encoded in the action for the probe Dq brane:
\be
S = - N_f T_{D_q} \int_{D_q} d^{q+1} x  \, e^{-\phi}\sqrt{-\det(g + 2\pi \alpha' F)} 
+ {WZ}\, \label{borninfeld}\, .
\ee
The second term on the r.h.s. stands for the Wess-Zumino term which will not make any contribution 
to the equations of motion for the background and the fluctuations (see Appendix \ref{apWZ}). $T_{D_q} = 1/((2\pi l_s)^q g_s l_s)$ is the Dq-brane tension,  $g_s$ is the string  coupling constant and $g_{\mu\nu}$ is the pullback metric induced by the relevant background. 
As for the background, we will be concerned with the near horizon limit of a stack of non-extremal Dp-branes. The general
form for any $p$ is given by
\beqa
ds^2 &=& H^{-1/2}(- f dt^2 + d\vec x^2) + H^{1/2}\left(\frac{d\rho^2}{f} + \rho^2 d\Omega^2_{8-p}\right)\, ,
\nonumber\\
e^\Phi &=& H^{\frac{3-p}{4}}~;~~~C_{01...p} = H^{-1} \, ,
\eeqa
where $\vec x = (x^1,...,x^p)$ and
\be
H(\rho) = \left( \frac{L}{\rho}\right)^{7-p}~~~;~~~f(\rho) = 1- \left(\frac{\rho_0}{\rho}\right)^{7-p} \, .
\ee
The probe Dq-branes wrap an $n-$sphere in the directions transverse to the Dp-branes, so it is convenient to write the metric on $S^{8-p}$ in adapted coordinates,
\be
d\Omega_{8-p}^2 = d\theta^2 + \sin^2\theta \,  d\Omega_n^2 + \cos^2\theta \, d\Omega_{7-p-n}^2\, .
\ee
Setting $\psi = \cos\theta$ the classical Dq-brane embedding may be specified by a dependence $\psi = \psi(\rho)$.
On the probe brane, a $U(1)$ gauge field can be switched on that will also depend only on the radial coordinate, $A_{\mu}(\rho)$.
We shall make use of the dimensionless radial coordinate $u$, related to $\rho$ by
\be
u = \left(\frac{\rho_0}{\rho}\right)^\frac{7-p}{2}\, ,
\ee
in terms of which $f (u)= 1- u^2$ and the horizon lies at $u=1$. The Hawking temperature is given by
\be
T = \frac{7-p}{4\pi L}\left(\frac{\rho_0}{L}\right)^{\frac{5-p}{2}} \, .
\ee

\subsection{The D3/D7 system}
From this point on we will specialize to the case of D7-brane probes in a black D3-brane geometry. The D3/D7 intersection is summarized in the following array
\be
\begin{array}{lcccccccccc}
          & 0 & 1 & 2 & 3 & 4 & 5 & 6 & 7 & 8 & 9 \\
  D3: & \times &  \times &  \times &  \times &    &    &    &    &     &     \\
  D7: &   \times &  \times &  \times & \times &  \times &  \times &  \times &  \times  &     &
\end{array}
\ee
and the bulk metric reads
\be
ds^2 = \frac{(\pi T L)^2}{u}(- f dx_0^2 + d\vec x^2) + \frac{L^2}{4u^2}\frac{du^2}{f} + L^2d\Omega_5^2 ~ \, ,
\ee
where
\be
L^4 = 4\pi g_s N_c l_s^4 \, .
\ee
Specifiying the D7-brane embedding through $\psi = \psi(u)$ the induced metric takes the form
\be
ds_{D7}^2 =
\frac{(\pi T L)^2}{u}(- f dt^2 + d\vec x^2) + \frac{L^2(1-\psi^2+4u^2f\psi'^2)}{4u^2f(1-\psi^2)}du^2 + L^2(1-\psi^2)d\Omega_3^2\, ,
\ee
with the D7-brane wrapping an $S_3\subset S_5$. The generalization of the previous setup for finite baryon density was investigated in \cite{hep-th/0611099}.   The relevant bulk degree of freedom dual to the baryon chemical potential  is the $A_{0}$  component of a $U(1)$ gauge field on the worldvolume of the D7-brane.
The background profiles for $\psi(u)$ and $A_{0}(u)$ are obtained by solving the Euler-Lagrange equations of the Born-Infeld lagrangian
\be
{\cal L} =  - N_f T_{D_7}    \sqrt{-\det(g + 2\pi \alpha' F)}\, .
\ee
The gauge field $A_{0}(u)$ obeys a conservation equation owing to the fact that it enters the action purely through its derivatives, \be
\partial_u
\left(
\frac{\tilde\psi^4 A_{0}'}{\sqrt{T^2L^4(\tilde\psi^2 +4u^2f\psi'^2) - \alpha'^2 16u^3  \tilde\psi^2 A_{0}'^2}}
\right) = 0\, ,
\label{profileA}
\ee 
where    ${\tilde\psi}(u)=\sqrt{1-\psi(u)^2} = \sin\theta(u)$. Asymptotically in the UV region $u\to 0$
we will show that $\lim_{u\to0}\psi = 0$ (see eq. (\ref{eq.UVpsi}) below) and therefore this equation reduces to $\d_u^2 A_{0}(u)=0$ which has the solution
\be
A_{0}(u) = \mu - a u + ...\, .
\ee
By means of the holographic dictionary \cite{arXiv:0705.3870}  $\mu$ is proportional to the chemical potential for the baryon number density, and $a$ is proportional to the baryon number density itself.
Equation (\ref{profileA}) implies the existence of a constant of motion, $\Dis$, which we normalize
as follows
\be
\Dis =
\frac{-4\alpha' \tilde\psi^4 A_{0}'}{\sqrt{T^2L^4(\tilde\psi^2 +4u^2f\psi'^2) - \alpha'^2 16u^3 \tilde\psi^2A_{0}'^2}}\, .
\ee
Evaluating this constant of motion at $u=0$ implies that
\be
\Dis = \frac{4\alpha'}{TL^2} a\, .
\ee
In terms of $\Dis$   the field $A_{0}$ can be expressed as
\be
A_{0}'(u) =- \frac{L^2 T}{4\alpha'}
\frac{\Dis \sqrt{\tilde\psi^2 + 4u^2f\psi'^2}}{\sqrt{\tilde\psi^2(\tilde\psi^6+\Dis^2 u^3)}}\, .
\ee
Following the discussion in \cite{hep-th/0611099} we can express the chemical potential as
\be
\mu=\frac{D\,TL^2}{4\alpha'}\int_{0}^1\frac{ \sqrt{\tilde\psi^2 + 4u^2f\psi'^2}}{\sqrt{\tilde\psi^2(\tilde\psi^6+\Dis^2 u^3)}}du\, ,
\ee
and we see that in the limit of vanishing baryon density $D\to0$ we obtain vanishing chemical potential (note that there is a region of the phase diagram for which this does not hold for sufficiently large quark mass \cite{hep-th/0611099,arXiv:0709.1225}).

The equation for $\psi(u)$ for a generic $Dp/Dq$ intersection is given in eq. (\ref{geneqpsi}), and, specializing to the D3/D7 case, gives  
\be
\partial_u \left(
\frac{4 f \tilde\psi ^2 \psi '\sqrt{\tilde\psi^6+D^2 u^3}}{u \sqrt{\tilde\psi^6(\tilde\psi^2+4 u^2 f
\psi{'}{^2})}}
\right) +
\frac{\psi \left(3\tilde\psi^4 +  4u^2 f \psi'^2 (2\tilde\psi^6-D^2u^3)\right)   }{u^3\sqrt{ \tilde\psi^6(\tilde\psi^6 + D^2 u^3)(\tilde\psi^2+4 u^2 f
\psi{'}{^2})}} =0\, .
\label{profilepsi}
\ee
Close to the boundary this equation reads $\partial_u\left( 4\psi'/u \right) = - 3\psi/u^3$ and its solution behaves as
\be
\psi(u) \sim \frac{m}{\sqrt 2} u^{1/2} + \frac{c}{2\sqrt{2}} u^{3/2} + ...\, ,  \label{eq.UVpsi}
\ee
independent of the baryon density and where the constants $m$ and  $c$ parametrize respectively  the quark mass and something we loosely refer to as quark condensate \cite{hep-th/0311270,hep-th/0304032,hep-th/0602174,hep-th/0605261,hep-th/0605017},
\beqa
M_q &=& \frac{1}{2} \sqrt{\lambda} T m\, , \nonumber\\
\langle {\cal O}\rangle &=& -\frac{1}{8}\sqrt{\lambda} N_f N_c T^3 c\, ,
\eeqa
with $\lambda = g_{YM}^2 N_c = 2\pi g_s N_c$, the 't Hooft coupling. The operator ${\cal O}$ is a supersymmetric version of the
quark bilinear
\be
{\cal O} = \bar\Psi\Psi + \Phi^\dagger X\Phi + M_q \Phi^\dagger \Phi\, ,
\ee
with $X$ one of the adjoint scalars.  A precise definition can be found in \cite{hep-th/0611099}. 

Equation (\ref{profilepsi}) is a non-linear differential equation which cannot be solved analytically and therefore its integration is performed numerically  \cite{hep-th/0306018,hep-th/0605046}. For $\Dis=0$ the stable embeddings $\psi(u)$ fall into two categories, labelled ``Minkowski" and ``black hole". Additionally one can find metastable configurations corresponding to a supercooled phase.
For Minkowski embeddings the D7-brane never enters the black hole and we have a maximum value $u_{max}<1$ the probe branes can reach. In this case one has stable bound states identified with mesons whose spectrum manifests a mass gap and discretization \cite{arXiv:0711.4467}. This is seen in the spectral function as an infinite sum of $\delta$-functions.
For black hole embeddings, the D7-brane intersects the horizon, hence $u_{max}=1$. The branes develop an induced horizon and all the meson resonances become unstable  \cite{hep-th/0612169,hep-th/0701132}.

\begin{figure}[ht]
\begin{center}
\includegraphics[scale=1.4]{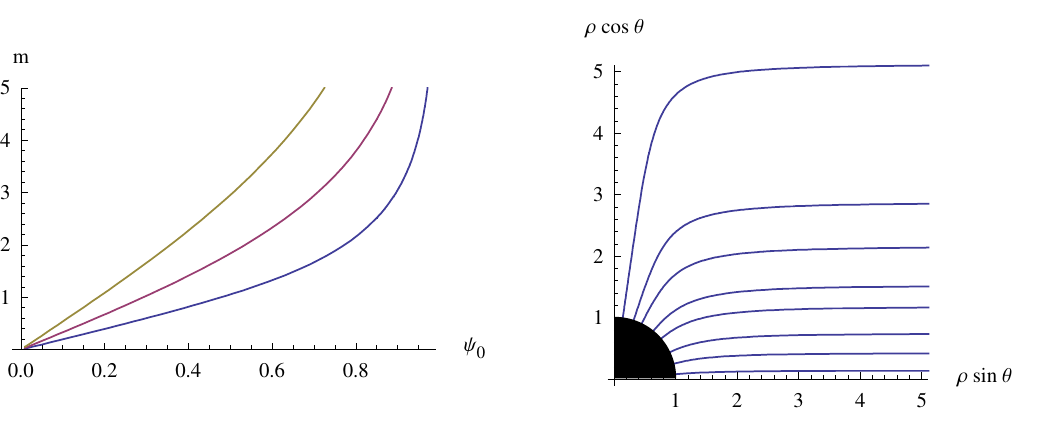}
\caption{\em \label{m1versusang1}
On the left-hand side we see the dependence of $m$ with $\psi_0$ for values of $\Dis=20,5$ and $1$ (from left to right). As soon as $\Dis\neq 0$, arbitrarily high values of $m$ have a black hole embedding with $\psi_0\in [0\, ,\, 1)$.
On the right-hand side we plot the  embeddings $\psi (u)$
for different values of $\psi_0$ ranging from  $0.1$ to $0.99$ and $D=1$.
}
\end{center}
\end{figure}

This situations changes drastically at finite baryon density.
One of the main discoveries in     \cite{hep-th/0611099} was that for any value  of $\Dis> 0$ there are only  black hole embeddings. 
In fact a regular series expansion around the horizon $u\sim 1$ that solves (\ref{profilepsi}) takes the form
\be
\psi(u) = \psi_0 - \frac{3}{8}\frac{\psi_0(1-\psi_0^2)^3}{(1-\psi_0^2)^3+D^2}(1-u) + {\cal O}(1-u)^2
\ee
and as we see, depends solely on the limiting value of the embedding profile at the horizon $\psi_0=\psi(u=1)$. Note that this solution automatically satisfies the orthogonality condition of  \cite{hep-th/0611099} due to the vanishing Jacobian on the horizon.
This expansion can be used to specify the boundary conditions in order to numerically integrate $\psi(u)$ out towards the boundary, from where we can read off $m$ and $c$ from eq. (\ref{eq.UVpsi}).
Plotting $m(\psi_0)$ in figure \ref{m1versusang1} we observe indeed that moving the 
value of $\psi_0$ in the range $[0 ,  1)$ covers the full range of  $m\in [0,  \infty)$.
\begin{figure}[ht]
\begin{center}
\includegraphics[scale=1.5]{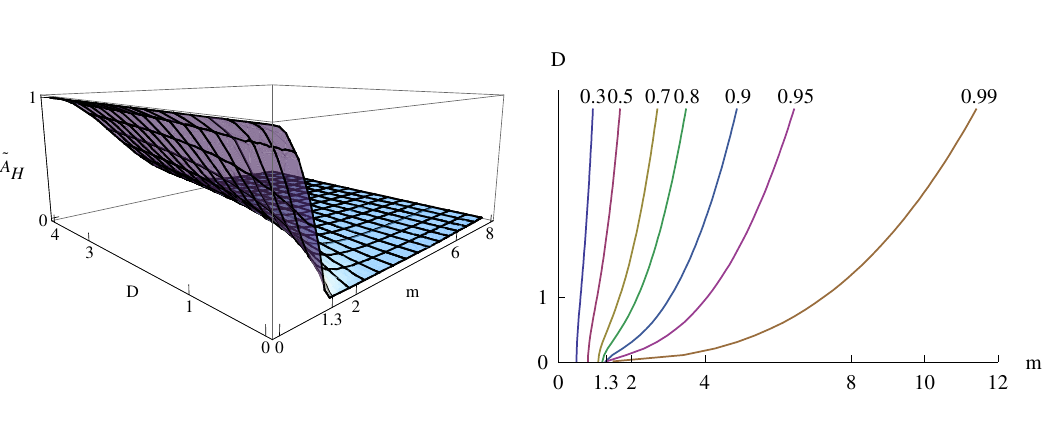}
\caption{\em \label{contourplot}
 The left plot shows the normalized horizon area as a function of the quark mass and the baryon number density. The right hand graph is a contour plot showing lines of equal induced horizon area, labelled by the corresponding value of $\psi_0$.
Hence we have smaller induced  horizons for larger values of $\psi_0$.
}
\end{center}
\end{figure}
The 3-area of the induced horizon (per unit  3-dimensional  Minkowski space volume) is controlled by  $\psi_0$
\be
A_H =2\pi^2(\pi T L^2)^3 (1-\psi_0^2)^{3/2}\, .
\ee
  In figure \ref{contourplot} we have plotted the curves of equal induced area on the  D7-brane.  
We expect this quantity to govern the rough shape of the peaks of the spectral function with larger widths for larger induced horizons. Close to $\Dis =0$ we see a qualitative change in behaviour  below $m_{crit}\sim 1.3$. This is roughly the critical mass below (above) which we have black hole (Minkowski) embeddings
  at $\Dis=0$. From the form of the iso-curves, we expect a maximum effect of $\Dis$ on physical quantities for $m$ in the range
  $\sim(1.3, 4)$. For $m$ outside this range the change in induced horizon area is small for a given increment in $D$.

\section{Fluctuations}\label{fluctuations}

We will consider perturbations of the world-volume fields that may depend only upon the
RG flow coordinate $u$ and the Minkowski coordinates $x^0,x^1$. In other words, we shall consider
only fluctuating fields that are independent of the internal coordinates wrapping the $S^3$.
\beqa
\psi(u,x) &\to &\, \psi(u)\,  + \, \epsilon \,  e^{-i(\omega x^0 - q x^1)}\Psi(u) \, ,\nonumber\\
A_{\mu}(u,x) &\to & A_{\mu}(u) + \epsilon \,  e^{-i(\omega x^0 - q x^1)}\dA_\mu(u)\, .
\eeqa
With this we can expand the DBI lagrangian in powers of $\epsilon$
\be
{\cal L} = {\cal L}_0 + \epsilon {\cal L}_1 + \epsilon^2 {\cal L}_2 + ...\, .
\ee
Upon imposing the equations of motion for the background fields,   ${\cal L}_1$ vanishes and
the linearized equations for the perturbations can be  derived from the quadratic piece.
The equations of motion can be found in appendix \ref{eqomo}.
For $q\neq 0$   the fields $\dA_0,\dA_1$ and $\Psi$ must solve a coupled system of differential equations
and cannot be set to zero independently. This   coupling only happens for $q\neq 0$ and vanishes in the limit of zero $\Dis$. On the other hand, the transverse excitations
$\dA_\perp = \dA_{2,3}$ do decouple. We will analyze the two sets of fields independently in the following section.

It is important to note our definition of the various degrees of freedom. Although the scalar and the longitudinal mode of the gauge field do mix via their equations of motion, we will show that the spectral functions decouple in the UV (that is there is a vanishing two-point function between the operators sourced by the scalar and the longitudinal vector field). This means that although there is IR operator mixing, there is no ambiguity in the definition of the vector and scalar modes on the UV boundary.
\subsection{Longitudinal and Scalar excitations}

This set of degrees of freedom is rather cumbersome 
to analyse because their equations of motion are coupled at first
order. 
 We have the freedom to make a gauge choice which leaves us
with a constraint, plus three linearly independent equations
relating $\dA_0$, $\dA_1$ and $\Psi$ (see eqs. (\ref{eomlong1} - \ref{eomscal})). By writing linear combinations of the constraint with the equations of motion for $\dA_0$, $\dA_1$ and $\Psi$ we
are able to find a single propagating gauge invariant combination
given by the longitudinal electric field component: \be \dZ=q\dA_0+\omega \dA_1\, . \ee This field is still coupled
to the scalar excitation $\Psi$ and thus the equations of motion for
these two degrees of freedom must be solved simultaneously (see eqs. (\ref{gauge1},\ref{gauge2})).
In order to solve the coupled equations of motion we must investigate
the boundary conditions  for the coupled system. For retarded Green's functions, we select incoming wave boundary conditions  on
the black hole horizon. The analysis of the Frobenius expansion is given in appendix \ref{apbound}. The selection of incoming wave boundary conditions   is possible and from it we can 
numerically integrate the coupled system of equations simultaneously to the boundary to obtain
the Fourier bulk modes $\Psi_k(u)$ and $\dZ_k(u)$ with $k=(\omega,q,0,0)$.
At this stage we must write the boundary action in terms of the $\Psi_k(u)$ and $\dZ_k(u)$ degrees of freedom.
This is  given by an expression of the form:

\begin{eqnarray}
S_{B}[\dZ,\Psi; u]&=&- {\cal N}\int \frac{d^4k}{(2\pi)^4} \left(F_{\dZ\dZ} \dZ'_k \dZ_{-k} + F_{\dZ\Psi} (\dZ'_k \Psi_{-k} +  \Psi'_k \dZ_{-k})
+F_{\Psi\Psi} \Psi'_k \Psi_{-k}  +...\right)\, ,
\label{longboundaction}
\end{eqnarray}
where the ellipses stand for non-derivative terms that will not contribute to the imaginary part of the spectral functions and 
\begin{equation}
{\cal N}=N_f T_{D_7} (2\pi^2)(2\pi \alpha')^2(\pi TL^2)^2 = \frac{N_f N_c T^2}{4}\, .
\end{equation}
The coefficients, $F_{IJ}$, are functions of the embedding solution, $\Dis$, $\omega$, $q$ and $u$ which are best given in terms of the following functions
\beqa
g(u) &=&  -(\tilde{\psi}(u)^6 -\Dis^2 u^3)  \, ,\\
h(u) &=& 
\frac{g(u)}{\tilde\psi(u)^{10}\left(\psi(u)^2 - 4 u^2 f(u)\tilde\psi'(u)^2-1\right)}
 \, .\label{deffgh}
\eeqa
We find, in terms of the   usual dimensionless ratios
$\wn = \omega/2\pi T$ and $\qn = q/2\pi T$,

\beqa
F_{\dZ\dZ}(k,u)&=&  \frac{  \tilde\psi^6  f g\sqrt{h}}{\qn^2 f \tilde\psi^6 + \wn^2 g}\frac{1}{(2\pi T)^2}~~
\stackrel{u \to 0}{\longrightarrow} ~~\frac{1}{\omega^2 - q^2}+ ...\, ,
\nonumber\\
F_{\dZ\Psi}(k,u) &=&   -\frac{4\qn \Dis \,  u^2 f^2 h \psi'\tilde{\psi}^{10}}{\qn^2 f\tilde\psi^6 + \wn^2 g}
\left(\frac{\pi TL^2}{2\pi \alpha'}\right) \frac{1}{2\pi T}
~~
\stackrel{u \to 0}{\longrightarrow} ~~2 mD u^{3/2}\frac{q}{\omega^2-q^2}   
 \left( \frac{\pi T L^2}{2\pi \alpha'}\right)+ ...\, ,
\nonumber\\
F_{\Psi\Psi}(k,u)&=& -\frac{f \tilde\psi^{14} h^{3/2}(\qn^2(f\tilde\psi^8-4\Dis^2 f^2\psi'^2 u^5)+ \wn^2 \tilde\psi^2 g)}{u g(\qn^2 f\tilde\psi^6 + \wn^2 g)}
\left(\frac{\pi T L^2}{2\pi \alpha'}\right)^2\, \nonumber\\
&& \stackrel{u \to 0}{\longrightarrow} ~~ \frac{1}{u}\left( \frac{\pi T L^2}{2\pi \alpha'}\right)^2+ ...\, .
\eeqa
From these expressions it is clear that  the coupling between operators sourced by $\dZ$ and $\Psi$ in the UV occurs only for  non-zero $q$ and $\Dis$.   

One can show that, on shell, the imaginary part of  the boundary action (\ref{longboundaction}) is independent of $u$. Despite the fact that this is of little help in  computing the individual components of the retarded two-point function $G^R_{IJ}(k)$, we have used this as a quality check of our numerical integration. In fact, obtaining a constant value
up to one part in $10^5$  supports our  confidence in the accuracy of the solutions.

In order to calculate the spectral function, we must take
derivatives with respect to the boundary values of the fields which
act as sources for the operators of interest. However, with
the $\Psi$ degree of freedom in its current form, this is not the
source for the quark-bilinear operator which we are interested in.
By looking at the solution of the field on the boundary (eq. (\ref{eq.UVpsi}))
we see that the field with the appropriate scaling to source a
$\bar{q}q$ operator (and its gaugino and scalar counterparts) in the UV is
$\tilde{\Psi}(u)=u^{-\frac{1}{2}}\Psi(u)$. 
Changing variables to $\tilde{\Psi}$ we can calculate the retarded correlator in the $\tilde\Psi, {\cal Z}$ sector which is given by $G^R_{IJ}(k), \, I,J = {\cal Z},\tilde \Psi$:
\beqa
G^R_{IJ}(k)&=&-{\cal N} \left.\left(
\begin{array}{cc} \displaystyle
2F_{\dZ\dZ}(k,u)  \left(\frac{\dZ_k'(u)}{\dZ_k(u)}\right) &
 \displaystyle
  F_{\dZ\Psi}(k,u) u^\frac{1}{2}\left(\frac{\dZ'_k(u)}{\dZ_k(u)}-\frac{\tilde{\Psi} '_{-k}(u)}{\tilde{\Psi}_{-k}
   (u)}\right)\\
    \displaystyle
- F_{\dZ\Psi}(k,u) u^\frac{1}{2}\left( \frac{\dZ'_k(u)}{\dZ_k(u)}-\frac{\tilde{\Psi}_{-k} '(u)}{{\tilde\Psi}_{-k}
   (u)}\right)&
    \displaystyle
    2u F_{\Psi\Psi}(k,u) 
  \left(\frac{{\tilde\Psi}_k '(u)}{{\tilde\Psi}_k
   (u)}\right)
\end{array}
\right)\right|_{u\rightarrow 0}\hspace{-0.7cm}
\nonumber\\
&& +\cdots\, ,
\label{retarcorr}
\eeqa
where the ellipses stand for real contact terms that will not contribute to the spectral function. The boundary limit 
$u\to 0$ corresponds to a definition of the dual theory in the UV. We see that in this limit the off-diagonal terms vanish and we obtain a diagonal matrix.  Conversely,  setting $u=\epsilon$ corresponds to a definition of the dual QFT as a Wilsonian effective field theory defined at a finite energy scale. In this case the presence of $\Dis$  will  induce  in the RG flow  a mixing of the corresponding quantum operators.

\subsection{Transverse excitations}

Transverse excitations  are decoupled and therefore simpler to
deal with in the presence of $\Dis$.  Again, computing the retarded Green
function involves the evaluation of the boundary action on relevant solutions
for the bulk perturbations. 
The equation of motion for these perturbations can be found in equation (\ref{eomperp}). 
 Close to the horizon the incoming transverse mode has the form:
\beqa 
\dA_{\perp,k}(u)\sim (1-u)^{-i\frac{\wn}{2}} a_k(u)\, ,
\eeqa 
with $k=(\omega,q,0,0)$ and $a_k(u)$ an analytic function at $u=1$. The solution
for $\dA_{\perp,k}(u)$ will depend parametrically on $m$ and 
$\Dis$ (both explicitly and through $\psi_0(m,\Dis)$), $\omega$ and $q$.
These solutions are to be inserted into the boundary action which is given by
\be
S_{B}[A_\perp, u] =-{\cal N} \int \frac{d^4k}{(2\pi)^4} \, F_{\perp}(u)\, \dA'_{\perp,k}(u)\dA_{\perp,-k}(u)\, ,
\ee
where $F_{\perp}(k,u) = f\tilde \psi^6\sqrt{h}$, with $h(u)$ as in (\ref{deffgh}) of the previous section.
An analysis of the differential equation shows that close to the boundary the exponents of the Frobenius expansion are $\Delta = 0,1$ as usual for a bulk gauge field. In this case, the usual prescription
for calculating Minkowskian 2-point functions is implemented \cite{hep-th/0205051}

\beqa \label{trangreen}
G^R_{\perp}(k)= -  2 {\cal N} \lim_{u\to 0 }F_\perp(k,u)
\left[\frac{\dA_{\perp,k}'(u)}{\dA_{\perp,k}(u)}\right]
\, .\eeqa

In section \ref{graphressec} we plot the results of this analysis. When comparison is possible, we find agreement with  the results of \cite{arXiv:0710.0334,arXiv:0804.2168}.


\section{Graphical results} \label{graphressec}

In this section we give an account of the results  obtained by numerical integration of the
equations of motion. In all cases we calculate the spectral function by using the definitions given in equations (\ref{spectdef}), (\ref{retarcorr}) and  (\ref{trangreen}). The equations of motion are defined in terms of the parameters $\qn$, $\wn$, $\Dis$ and implicitly through the mass $m$. In the following we will explore what we believe to be the most enlightening areas of this large parameter space. In various limits, the numerical results can be seen to coincide with the analytical behaviour, derived in appendix \ref{asymptote}.

\subsection{Vector modes}
\subsubsection{Lightlike momenta $\qn = \wn$}

For the study of spectral functions at lightlike momenta, the longitudinal excitations vanish and we must simply solve for the transverse degrees of freedom. Spectral functions at lightlike momenta have been calculated for $\Dis=0$ in \cite{arXiv:0709.2168} as they
are relevant for the computation of  the emission of thermal photons from the plasma.
\begin{figure}[ht]
\includegraphics[scale=1.8]{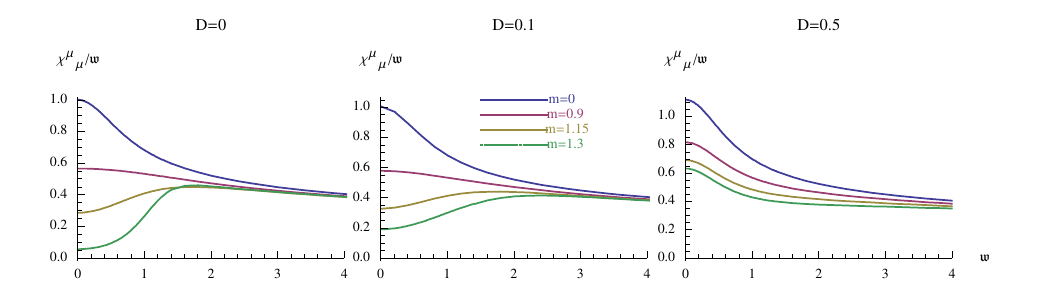}
\caption{
\em \label{lightlikeDis}
In this figure we plot  the quotient of the spectral function over the frecuency $\chi^\mu{_\mu}(\wn)/\wn$  for lightlike momenta
$\wn=\qn$ and different values of the constituent quark mass. 
Above   $\Dis \sim 0.3$ all curves become monotonically decreasing with $\wn$.
}
\end{figure}
 \begin{figure}[ht]
\begin{center}
\includegraphics[scale=1.8]{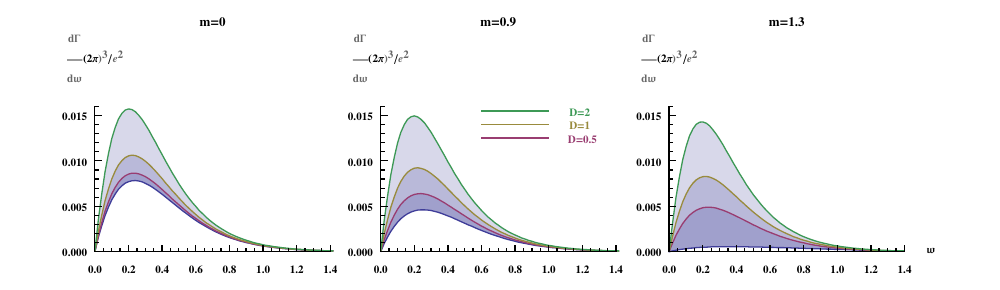}
\caption{
 \em  \label{photonrateDdep}
Illustration of the increase in the photoproduction rate with $D$ for masses below $m_{crit}$. }
\end{center}
\end{figure}
 For lightlike momenta the longitudinal contribution to the spectral function vanishes.  In figure \ref{lightlikeDis} we have plotted\footnote{In all the plots we normalize the spectral function in units of $\frac{N_fN_cT^2}{2}$ and the scalar mode in units of $\frac{N_f N_c T^4 L^2}{8\alpha'^2}$.} the spectral function for different masses and baryon density. The left most plot coincides  with the results in \cite{arXiv:0709.2168} and the others show the influence of $\Dis$ on the set of curves labelled by different masses. 

As a general rule we find an enhancement of the spectral function for low values of the momentum. 
The top curve corresponds to massless quarks, and its behaviour at the origin is related to the electrical conductivity. We observe an increase with $\Dis$ that will be investigated below.
In fact,  for sufficiently high values of $\Dis(\gtrsim 0.3)$ all the curves for different masses become monotonically decreasing functions of $\omega$.
 A peculiar feature at  $\Dis=0$ is the crossing of the curve $m=1.3$
with curves of slightly lower mass. This, translated into emission rates suggests that plasmas made of heavier quarks would shine brighter beyond $\wn\sim 1.4$. We see that this feature disappears already for low values of the 
baryon number and $\chi^\mu{_\mu}$ becomes a single valued function of $\omega$ and $m$.
The enhancement of the spectral function, translated into photoproduction rates exhibits a considerable increase as plotted in figure \ref{photonrateDdep}. The white area below the coloured curves corresponds to the $\Dis=0$ value. The enhancement is extremely large for masses around $m_{crit}$ (the mass of the quark at the phase transition point in the absence of baryon number $\sim 1.3$). This corresponds to an effective increase in brightness of the plasma of more than an order of magnitude!

\subsubsection{Large $\wn$ behaviour}
\begin{figure}[ht]
\begin{center}
\includegraphics[scale=1.7]{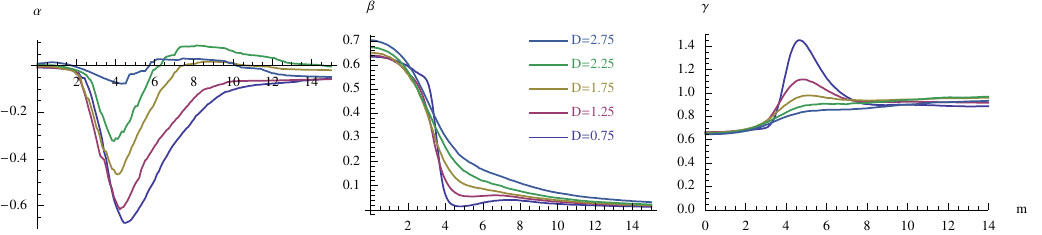}
\caption{
 \em  \label{largewbeh}
Asymptotic behaviour of $\chi^\mu{_\mu}$ as a function of the mass. We see that for large masses the
coefficients of the fit stabilize around a value of $\gamma\sim 1$. Note that numerical errors are large here due to the region of $\wn (\sim 70)$ in which the calculation is being performed.
}
\end{center}
\end{figure}
From the analysis in  \cite{hep-th/0607237} we know that  at large $\omega\gg1$  the spectral function for lightlike
momenta goes like $\sim \omega^{2/3}$ when $m=0$. In \cite{arXiv:0709.2168}, this behaviour seemed to persist for finite  mass  less than or equal to the critical one $m\leq 1.3$.
These are the only stable black hole embeddings for
vanishing $\Dis$.  To go to higher quark masses one must turn on $\Dis$ for stability.
 We have done a similar analysis and fitted our numerical results to a function of the form 
 $\chi^\mu{_\mu} = \alpha + \beta \wn^\gamma$ where the $\alpha, \beta$ and $\gamma$ are functions of $m$ and $\Dis$. The results are plotted in figure \ref{largewbeh}.
The universality of $\gamma\sim 2/3$ as observed in \cite{arXiv:0709.2168} corresponds to the
 the right most figure for values of $m\leq m_{crit}\sim 1.3$ where we see that all curves accumulate on an almost flat line.
For masses approximately in the range $m\in (1.3, 6)$ the parameters of the fit exhibit a strong dependence on the baryon number density. This is to be expected, as
these embeddings are the ones for which the induced horizon area changes most dramatically when $\Dis$ is switched on (see figure \ref{contourplot}). For large $m$ despite the fact that the embedding changes greatly in the UV, the change in the induced horizon area is negligible. For asymptotically large values of $m$ the parameters seem to  stabilize around some marginally $\Dis$ dependent values.

\subsubsection{Timelike momenta}

\begin{figure}[ht]
\begin{center}
\includegraphics[scale=1]{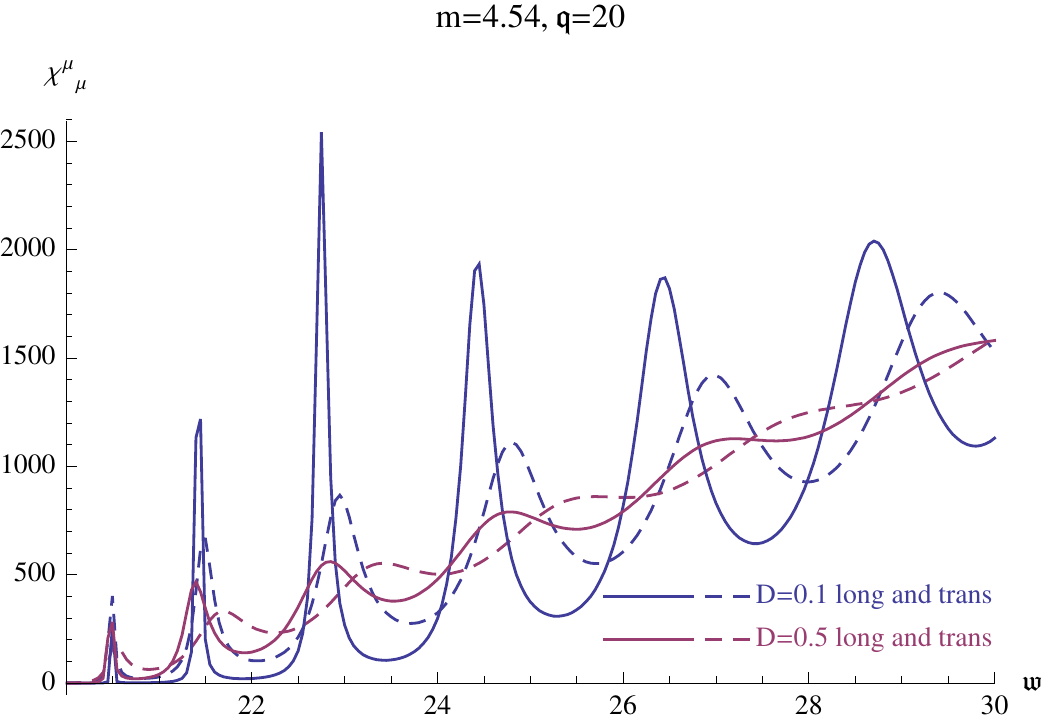}
\caption{
\em \label{PeaksDis20}
This plot shows the spectral function for longitudinal and transverse modes $\chi^{||}(\omega)$ and 
$\chi^{\perp}(\omega)$  at $\qn=20$ for two different values of  $\Dis=0.1,$ and $0.5$. All curves
oscillate around the zero temperature  value given in 
eq. (1.3).
}
\end{center}
\end{figure}

\begin{figure}[h]
\begin{center}
\includegraphics[scale=1.6]{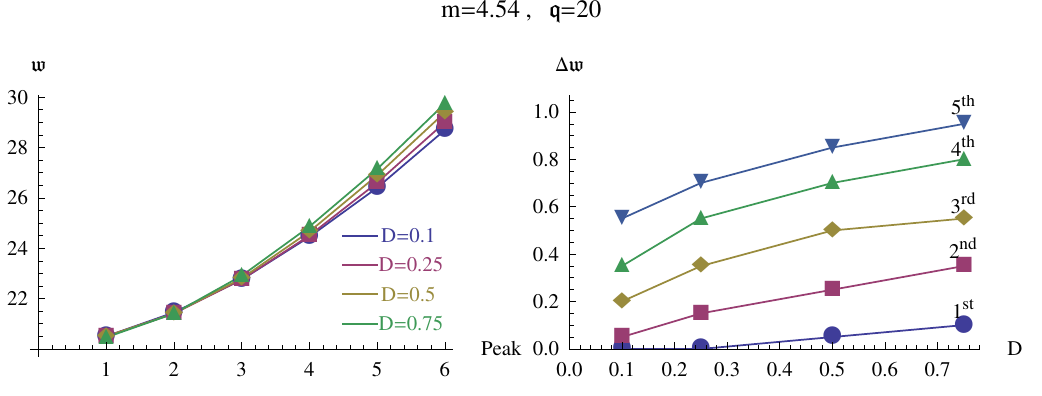}
\caption{
\em \label{peaksmove}
Left hand plot: position of the first peaks in the longitudinal component $\chi^{||}$.  Right hand side: difference between the positions of the $n^{th}$ peaks of $\chi^{||}$ and $\chi^\perp$.  All for different values of $\Dis = 0.1,0.25,0.5$ and $0.75$.
}
\end{center}
\end{figure}

For timelike momenta both $\chi^{||}$ and $\chi^{\perp}$ are nonzero and contribute to the spectral function. In figure \ref{PeaksDis20}  we have plotted both functions for several values of the baryon density, $\Dis=0.1$ and $0.5$, in the range where the peaks are clear. As a function of the baryon density, the peaks seem to decrease in amplitude  and increase in width monotonically. This corresponds to the poles moving further from the real axis  in the complex plane.  Concerning the spread among both components, the situation is not easily captured from the
 figure, therefore we have  tracked the movement of those peaks in figure \ref{peaksmove}. 

On the left we plot the position of the peaks in $\chi^{||}$,  and see a drift towards higher values
for increasing $\Dis$. On the right, we observe how the difference in peaks among both components
indeed builds up also with $\Dis$.

\begin{figure}[h]
\begin{center}
\includegraphics[scale=1.7]{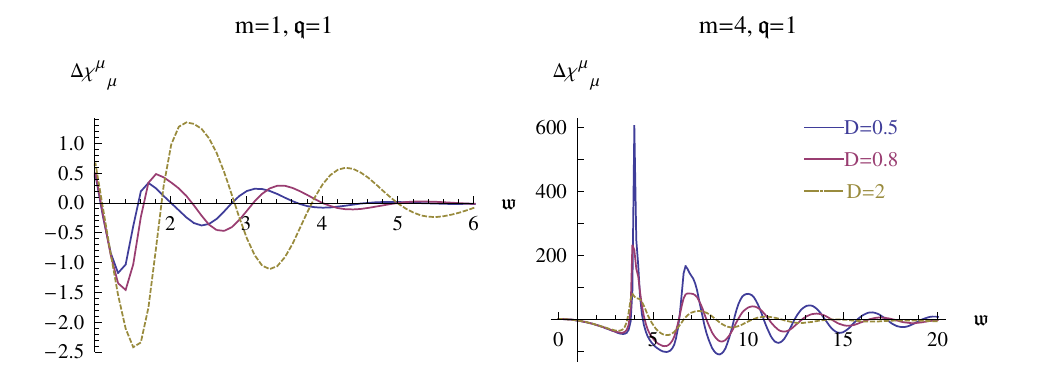}
\caption{
 \em  \label{mass1q10}
Finite $T$ contribution to the full spectral function $\Delta\chi(\wn,\qn=1)$ for different values of $\Dis$. }
\end{center}
\end{figure}

In figure \ref{mass1q10} we plot the difference in the full spectral function from the  $T=0$ result as a function of $\wn$
for $\qn=1$ and different values of the baryon density.
For values of $m$ below (above) $1.3$ the effect of the chemical potential is to enhance (suppress)  the height of the oscillations as a function of $\wn$. 
\begin{figure}[h]
\begin{center}
\includegraphics[scale=1]{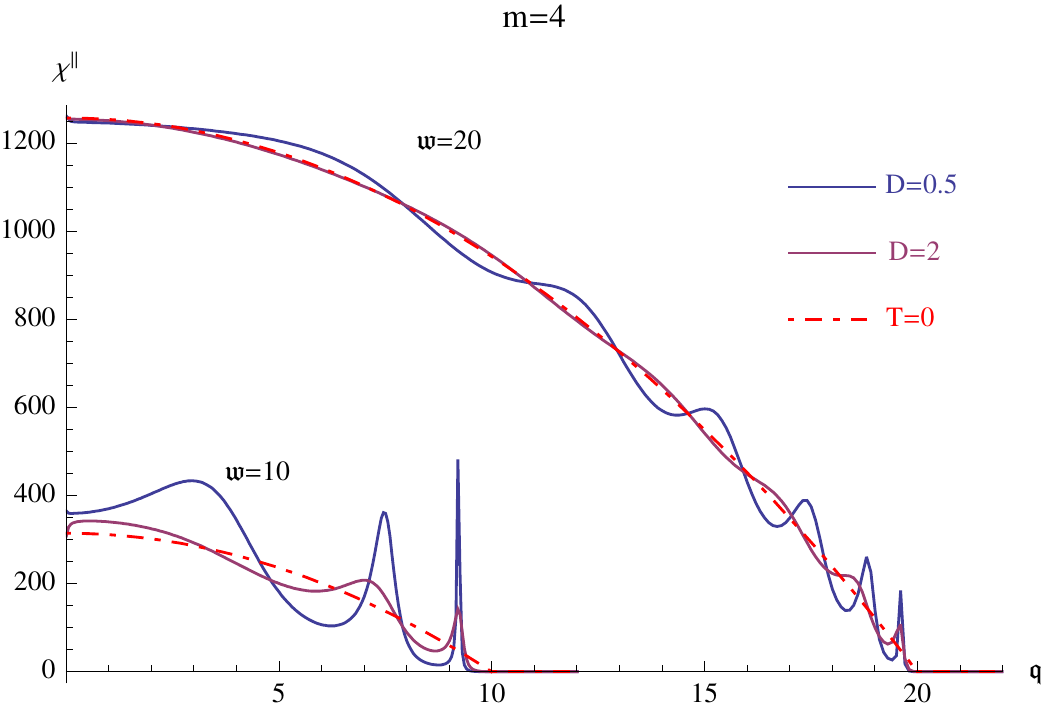}
\caption{
 \em  \label{mass4w1020q}
$\qn$ dependence of the longitudinal spectral function at varying $D$ and $\wn$. }
\end{center}
\end{figure}

\begin{figure}[ht]
\begin{center}
\includegraphics[scale=1.8]{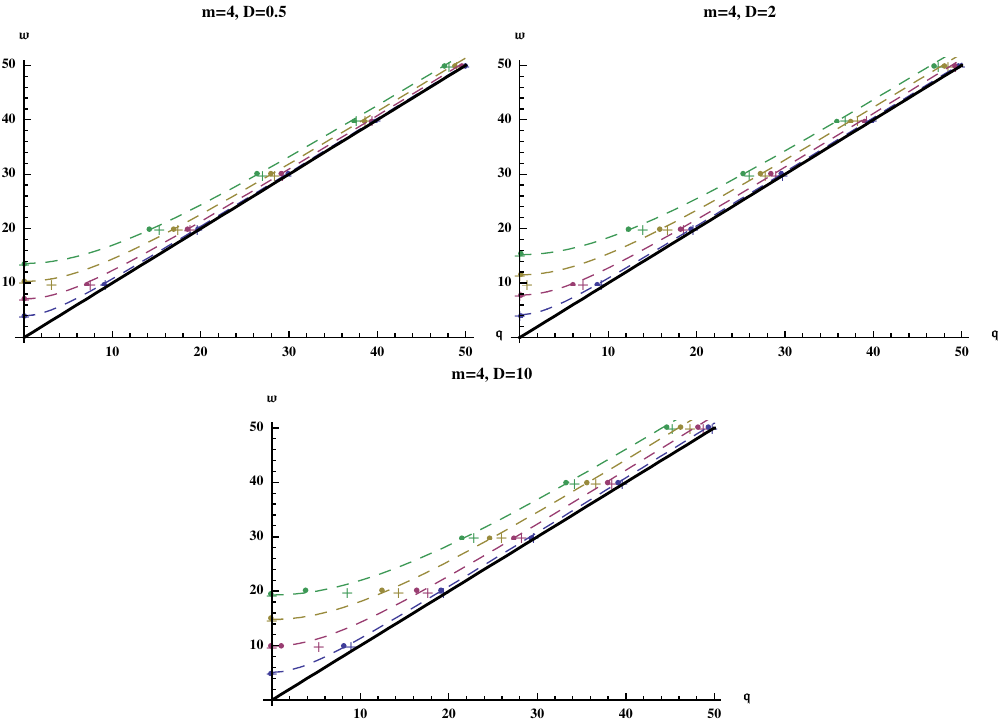}
\caption{
 \em  \label{limvelom4}
For the transverse spectral function, the first four peaks move in the $\wn$, $\qn$ plane
along  hyperbolae with different  limiting velocities, data points are marked with full circles. For the
longitudinal modes we do not have a nice parametrization, so we plot only the raw data, marked as crosses. The thick line is the light cone $\qn=\wn$.
 }
\end{center}
\end{figure}

Concerning the $\qn$ dependence,  figure \ref{mass4w1020q} shows the longitudinal spectral function, $\chi^{||}$, for two different values of $\wn=10$, and $20$.
For such high values we expect results very close to the zero temperature prediction as given in
eq. (\ref{zerotemp}) whose plot corresponds to the dashed line in the figure. 

We see peaks appearing in the  region close to the light cone ($\qn\sim\wn$). As we move away from the light cone (to smaller $\qn$) the peaks in the spectral function for different $\wn$ are in one-to-one correspondence. This could be seen by plotting the spectral function in the full $\wn-\qn$ plane.
As before, we must treat peaks in the longitudinal and transverse spectral functions separately.
We have plotted them in figure \ref{limvelom4}. Peaks in $\chi^\perp$ fit very well with a mass 
hyperbola of the form
\begin{equation}\label{masshyp}
\wn^2 - v_{\perp,n}^2 \qn^2 = M_n^2
\end{equation} 
where $M_n$ should correspond to the modes of the quasiparticle spectrum, and presumably coincides with the continuation of the mass spectrum  to the region where they become unstable (see the discussion in references  \cite{arXiv:0706.0162,arXiv:0803.0759}).
A careful analysis of the limiting velocity, $v_n$, of these unstable mesons was initiated in
\cite{arXiv:0712.0590} and recently pursued in  \cite{arXiv:0804.2168}. 
For $\Dis = 0.5,\, 2,\, 10$ our data are consistent with the fits in table (\ref{table1})\footnote{We fit the mass hyperbola eq. (\ref{masshyp}) by setting the value of $M_n$ equal to the peak position at $\qn=0$ and performing a $\chi^2$ fit analysis to the rest of the points.}.
In  \cite{arXiv:0804.2168} it was argued that the limiting velocity should be in any case $v_{\perp,n}=1$ for $\qn\to \infty$.
The interpolation to this behavior led the authors to speculate on  the existence of an intermediate region where the group velocity may become superluminal (see \cite{Fox:1970cu, Fox:1969us} for discussions on topics related to superluminal propagation).
\begin{equation}
\begin{tabular}{|c|c|c||c|c||c|c|}
\cline{2-7}
\multicolumn{1}{c}{} & \multicolumn{2}{|c||}{$D=0.5$} & \multicolumn{2}{c||}{$D=2$}& \multicolumn{2}{c|}{$D=10$} \\
\hline
$n$ & $M_{n}$ & $v_{\perp,n}$ & $M_{n}$ & $v_{\perp,n}$ & $M_{n}$ & $v_{\perp,n}$\cr
\hline
1 & 4.0 & 1.0006 & 4.2 & 1.0038 & 5.1 & 1.01251 \cr
2 & 7.1 & 1.0036 & 7.8 & 1.01086 & 9.9&1.02618\cr
3 & 10.3 & 1.0064 & 11.5 & 1.01762 & 14.8&1.03945\cr
4 & 13.6 & 1.0101  & 15.2 & 1.02422 & 19.3&1.04574\cr
\hline
\end{tabular}
\label{table1}
\end{equation}
 The departure from $v_{\perp,n}=1$ is certainly very small. However it displays a well defined monotonically increasing behaviour with $\Dis$ and $m$.  Due to rotational invariance for $\qn=0$ we know that the curves for the transverse and longitudinal peaks must intersect at zero momentum. However, unlike the transverse case, we were unable to fit the longitudinal peaks to a mass hyperbola.

\subsection{Scalar modes}
\begin{figure}[ht]
\begin{center}
\includegraphics[scale=1.8]{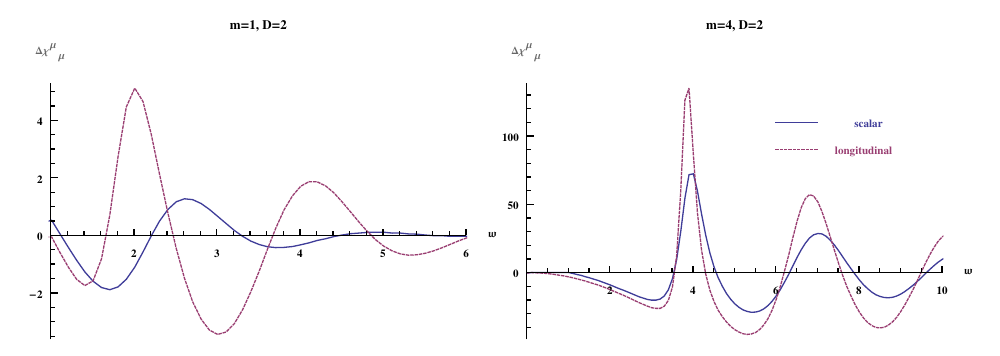}
\caption{
 \em  \label{scallongq1dis2}
For larger values of the quark mass $m$ the curves for the longitudinal spectral function and the scalar are more similar than for smaller values.
 }
\end{center}
\end{figure}
The spectral function for the scalar modes is obtained along the same lines as stated in the previous section from the $G^R_{\Psi\Psi}$ component in  eq. (\ref{retarcorr}). Again, they must be calculated by solving the full coupled equation, although the operator becomes decoupled in the UV.
In figure \ref{scallongq1dis2} we compare the longitudinal part of the spectral function and the scalar perturbation for two values of the quark mass $m$ at fixed $D$. For large masses we recover the $T=0$ result and therefore the peaks narrow and become degenerate for the different modes, while for low quark masses we see that the peaks broaden, as expected, and move away from the longitudinal modes  to which they are coupled in the IR. We see that the masses of the scalar quasinormal modes (the peaks in the spectral function) are higher than the equivalent longitudinal modes. The lifetimes are also shorter, given by a widening of the scalar peaks. Note that this same pattern is seen in a slightly different context in \cite{arXiv:0802.0775} for instance.


\section{Conductivity}\label{conductivitysec}

\begin{figure}[ht]
\begin{center}
\includegraphics[scale=1]{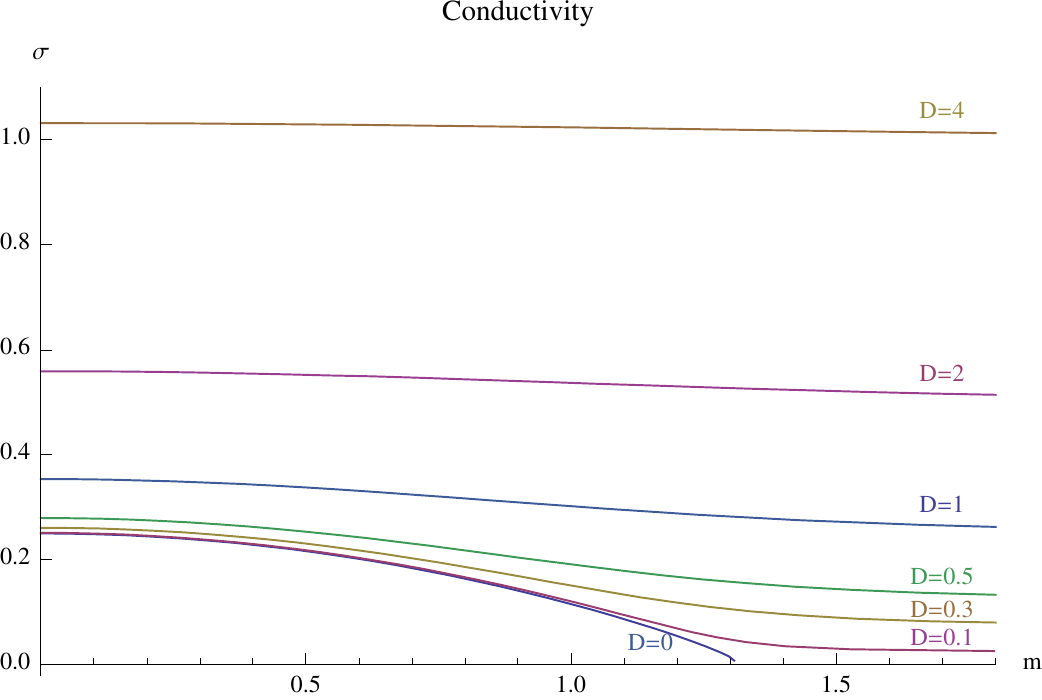}
\caption{\em \label{conductivity}
Conductivity, calculated numerically from the derivative of the spectral function as a function of the quark mass for various values of the baryon number density. The plot matches with the analytic expression given in 
eq, (5.2) upon setting $\varepsilon =0$.
}
\end{center}
\end{figure}

The electrical conductivity can be obtained from the zero-frequency slope of the spectral function for electromagnetic currents.
In \cite{arXiv:0709.2168} the conductivity of the plasma was calculated. We may do the same here, however we include the effects of a finite baryon density. This quantity may be calculated at null-momentum \cite{hep-th/0607237,arXiv:0709.2168} and therefore only the transverse spectral function contributes;
\begin{equation}
\sigma= \frac{e^2}{4}\lim_{k^0\to 0} \left. \frac{1}{k^0}\chi^\mu{_\mu}(k)\right\vert_{k^0= |{\bf k}|}\, .
\end{equation}
We show the results for this numerical calculation in figure \ref{conductivity}.
In \cite{arXiv:0705.3870} the conductivity of the plasma was obtained from Ohm's law using AdS/CFT techniques.
In the presence of an electric field $\varepsilon$ a current establishes satisfying Ohm's law $J= \sigma \varepsilon$. The 
electrical conductivity is given by the following expression
\footnote{Note that our definition of $\sigma$ and $\Dis$ differs from that in \cite{arXiv:0705.3870} by constant coefficients which are given in detail in appendix \ref{ap.KarchOB}.}:
\begin{equation}
\sigma =\frac{e^2}{4}\sqrt{\sqrt{\varepsilon^2+1}(1-\psi_0^2)^3+\frac{D^2}{\varepsilon^2+1}}\, , \label{karchbann}
\end{equation}
where  $\varepsilon$ the electric field strength, is zero in the present setup.
An insightful interpretation of this formula was given in the
aforementioned reference in terms of two sources of charge carriers
contributing
to the conductivity. One comes explicitly from the baryon charge
asymmetry, and is given by the $D^2$ contribution. The other comes
from charge carriers produced via pair creation in the plasma.  This effect  clearly depends upon
the mass of the fundamental charged quarks through the combination
$(1-\psi_0^2)$  and is  equal to  $1$ for massless quarks
and to $0$ for infinitely massive ones. This second effect is implicitly dependent on $\Dis$. Our numerical results plotted in
figure \ref{conductivity} match perfectly with the above analytic
expression
in the limit of vanishing electric field $\varepsilon \to 0$.

This result can also be derived from the
exact analytic expression for  the low frequency limit of the spectral function 
as given by (\ref{low-Spectral}).
The  calculation in \cite{arXiv:0705.3870}  is a macroscopic result. The
fact that this matches with our  derivation from microscopic linear
response theory is a highly non-trivial check of consistency since
both  explicit and implicit dependence on the baryon density are
correctly reproduced from the full spectral function.


\section{Conclusions}\label{conclusions}

Exploiting the AdS/CFT correspondence at finite temperature and baryon density we have considered a holographic dual of ${\cal N}=4$ super Yang Mills with quenched massive flavours (which break supersymmetry to ${\cal N}=2$). These flavours are constructed from a small number,  $N_f$, of  D7-branes embedded in the black hole geometry sourced by a stack of $N_c $ D3-branes. The addition of finite baryon number is modeled by turning on a vev for the worldvolume gauge field $A_0$ living on the D7-brane. Working at fixed baryon number gives an additional parameter $\Dis$ in addition to the constituent quark mass in units of the Hawking temperature. This opens up the possibility of studying the spectral functions for flavour bilinear operators in regimes where there is clearly recognizable structure. This structure is interpreted in terms of quasinormal modes of mesons which develop a finite lifetime, and therefore a broadening of their spectral functions at finite temperature. By looking at the spectral function of longitudinal and transverse modes as well as excitations of the scalar embedding coordinates on the D7-brane we have studied the behaviour of the poles in the spectral function as a function of $\Dis$ as well as spatial momentum. In particular, for finite $\Dis$ and $\qn$ we are forced to solve for the scalar and longitudinal modes as a coupled system of differential equations.

In general we find a splitting of the peaks between the spectral functions $\chi^\perp$ and $\chi^{||}$ which increases with both the baryon number and the spatial momentum. Plotting this effect for different quark masses, we find that the effect is most pronounced for values of $m$ in the range $\sim(1.3,4)$. This is where the induced horizon area changes most upon switching on $\Dis$. We have also investigated the dispersion relations $\wn (\qn)$, where $\wn$ is the position of the peaks in the spectral function. For the peaks in $\chi^\perp$ we find a very good fit of the data for the position of the peaks on a mass hyperbola of the form  $\wn=\sqrt{ v_n^2 \qn^2 + M_n^2}$, as discussed in \cite{arXiv:0804.2168}. For low masses, close to the critical mass, (as defined by the phase transition present in the absence of baryon density) and very low values of $\Dis$ (so that the peaks are still clearly seen) the values of $v_{\perp,n}\ll 1$ found in \cite{arXiv:0804.2168} are reproduced. 
For high enough values of the baryon density (and therefore high quark masses, in order to keep the position of the poles close to the real axis) we also find values of $v_{\perp,n}> 1$. For the longitudinal modes we do not know the correct function to fit. All of these issues should be confronted with the observation made in  \cite{arXiv:0804.2168}  that as $\qn\to \infty$ the zero temperature relativistic dispersion relation with $\wn =\qn$ should be recovered. The clarification of this issue demands the use of powerful techniques, in particular the analysis of the quasinormal modes. Certainly this will be a very challenging calculation for the coupled longitudinal and scalar modes, where the technique of converting the equation of motion to Schr\"odinger form becomes very difficult. We will come back to this issue in the future.

The most spectacular effects of the baryon density appear in the small $\wn$ region for lightlike momenta. This is seen in the set of figures \ref{lightlikeDis} where most of the change with $\Dis$ occurs for $\wn\lesssim 2$.  There are two main consequences of this. On one hand the baryon density is responsible for an increase in the photoproduction rate, for small quark masses by a very large percentage. On the other, the slope as a function of $\omega$ at the origin is directly related to the electrical conductivity. We have provided analytical and numerical evidence that the effect of baryon number is given by $\sigma\sim \sqrt{\left(1-\psi_0^2\right)^3 + D^2}$. This provides a highly nontrivial check of the results obtained in \cite{arXiv:0705.3870} for the electrical conductivity in a macroscopic setup.

It is clear that there is room for more research in this area, in particular a study of the limiting velocity and functional form of the dispersion relations for the longitudinal and scalar modes would be of great interest. In order to investigate this accurately, a large amount of computational time will be necessary. The study of the effects detailed in this paper may also be applied to various other finite temperature geometries. As mentioned above, a full calculation of the quasinormal modes will allow us to go to higher values of spatial momentum than are currently available by studying peak positions in the spectral functions.

\acknowledgments
We would like to thank warmly Alfonso V. Ramallo for his encouragement and always insightful comments.
Also it is a pleasure to thank  Daniel Are\'an, Nestor Armesto, Johanna Erdmenger, Nick Evans, Johannes Gro\ss e, Karl Landsteiner, Larry D. MacLerran, Cristina Manuel, David Mateos, Sergio Montero and Andrei Starinets for correspondence and discussions.
This  work was supported in part by MEC and  FEDER  under grant
FPA2005-00188,  by the Spanish Consolider-Ingenio 2010 Programme CPAN (CSD2007-00042), by Xunta de Galicia (Conselleria de Educacion and grant PGIDIT06PXIB206185PR)
and by  the EC Commission under  grant MRTN-CT-2004-005104.  JT has been supported by MEC of Spain under a grant of the FPU program. JS has been supported by the Juan de la Cierva program.


\appendix

\section{Equations of motion \label{eqomo}}

In this section we shall give more details about the equations we have been dealing with in the
text. We will try to keep quite general and provide formulae that encompass general $Dp/Dq$
intersections. The background metric will be given by a type II supergravity solution, of the following generic form
\be
ds^2_{II} =G_{MN}(X) dX^M dX^N~~,~~~~~~~~~~~~M,N= 0,...,9 \, .
\ee
We shall consider 
the embedding  in the following gauge: separate $X^M$ into two groups $X^a,\, a=0,1,...,q$ and
$X^I,\, I=q+1,...,9$.
\be 
X^M ~ \to ~ \left\{
\begin{array}{rcl}
X^a &=& \xi^a = \{ x^0,x^i,u,\theta^k \}~~~~~~~~(a=0,1,...,q)  \\
 X^I &=& Z^I(\xi) = \{ \psi,\varphi^j\} ~~~~~~~~~~(I=q+1,...,9) \rule{0mm}{5mm}
 \end{array}
 \right\}\, .
\ee 
Then the pullback metric and the dilaton are 10 dimensional bulk fields in origin. They are
evaluated on the $Dq$ brane
\be
\phi = \phi(\xi, Z(\xi )) ~~~~;~~~~
g_{ab} =  G_{ab}(\xi, Z(\xi)) +G_{IJ}(\xi,Z(\xi))\frac{\d Z^I}{\d\xi^a} \frac{\d Z^J}{\d\xi^b} \, ,
\ee 
where we have assumed that the background metric 
has no mixed terms $G_{a I} = 0$. On the other hand, the gauge fields are purely world-volume fields,
hence
\be
F_{ab}(\xi) = \d_aA_b (\xi)- \d_bA_a(\xi)\, .
\ee
Let us consider a set of perturbations of the form 
\be 
Z^I(\xi) = z^{(0)I}(\xi) +\epsilon z^{(1)I}(\xi) ~~~~; ~~~~~
A_a(\xi) =  A_a^{(0)}(\xi) + \epsilon A_a^{(1)}(\xi)\, ,
\ee
and expand up to second order in $\epsilon$. Naming $g_{ab}$ the pullback metric we have
\be
\phi =  \phi^{(0)} +\epsilon \phi^{(1)} +\epsilon^2 \phi^{(2)}  ~~~~; ~~~~~~
g_{ab} =  g^{(0)}_{ab} + \epsilon g_{ab}^{(1)} + \epsilon^2 g_{ab}^{(2)} ~~~;~~~~~
F_{ab} = F_{ab}^{(0)} + \epsilon F_{ab}^{(1)}  \, .
\label{expfield}
\ee
For example
\beqa
\phi^{(1)} &=& \d_I\phi^{(0)}z^{(1)I} ~~~~;~~~~~~
\phi^{(2)}  = \med \d_I \d_J \phi^{(0)} z^{(1)J}z^{(1)J} \, ,\nonumber\\
g_{ab}^{(0)} &=& G_{ab}(\xi, z^I(\xi))\, , \nonumber\\
g_{ab}^{(1)} &=& \left(  G_{ab,K}+  G_{IJ,K} z^{(0)I}_{,a}  z^{(0)J}_{,b}\right)\, z^{(1)K}
+   2G_{IJ} z^{(0)I}_{,(a}  z^{(1)J}_{,b)}\, ,
\nonumber\\
g_{ab}^{(2)} &=& \med \left(  G_{ab,KL}+  G_{IJ,KL}  z^{(0)I}_{,a}  z^{J(0)}_{,b} \right)\, z^{(1)K} z^{(1)L}
+  G_{IJ}  z^{(1)I}_{,a}   z^{(1)J}_{,b} + 
2 G_{IJ,K}z^{(1)K} z^{(0)I}_{,(a}  z^{(1)J}_{,b)}\, , \nonumber
\eeqa
We want to expand  the DBI lagrangian 
\be
{\cal L} = e^{-\phi} \sqrt{|\det( g + F)|}  \, ,
\ee
in powers of $\epsilon$. Note that the absence of the Wess-Zumino term is detailed in appendix \ref{apWZ}. Defining 
\be
\gamma_{ab} \equiv {\cal G}^{(0)}_{ab} = g^{(0)}_{ab} + F^{(0)}_{ab}~~~;~~~{\cal G}^{(1)}_{ab} = g^{(1)}_{ab} + F^{(1)}_{ab}~~~;~~~{\cal G}^{(2)}_{ab} = g^{(2)}_{ab}\, ,
\ee
as well as  
$$
\Sigma^{(i)}{^a}{_b} \equiv \gamma{^{ac}} {{\cal G}^{(i)}} _{cb}\, ,
$$  
the lagrangian up to  ${\cal O}(\epsilon^2)$ is
\beqa
{\cal L} =e^{-\phi}\sqrt{|\det{\cal G}|} &=& {\cal L}_0 + \epsilon{\cal L}_1+\epsilon^2 {\cal L}_2\, ,
\eeqa
with $\gamma = - \det \gamma_{ab}$ we have
\beqa
{\cal L}_0 &=& e^{-\phi^{(0)}}\sqrt{\gamma}\, , \nonumber\\
{\cal L}_1 &=& 
e^{-\phi^{(0)}}\sqrt{\gamma}  
\left( - \phi^{(1)} + \med \tr\Sigma^{(1)} ) \right)  \, ,\nonumber\\
{\cal L}_2 &=& e^{-\phi^{(0)}} \sqrt{ \gamma}  \left(
   \med\tr\Sigma^{(2)}  - \frac{1}{4}\tr \Sigma^{(1)2} + \frac{1}{8}(\tr\Sigma^{(1)})^2  
- \frac{1}{2} \phi^{(1)}\tr\Sigma^{(1)}+\frac{1}{2}\phi^{(1)2}- \phi^{(2)}\right)\, .
\eeqa
Let us now focus on a particular family of bulk metrics adapted to the setting of 
a $Dp/Dq$ intersection, with  $q = p+n+1$.
\beqa
ds^2_{II,B} &=& G_{00}(u) dx_0^2 + G_{ii}(u) d\vec x^2_p + G_{uu}(u) du^2 + 
G_{\theta\theta}(u,\psi)d\Omega_n^2 + G_{\psi\psi}(u,\psi)d\psi^2 + G_{\varphi\varphi}(u,\psi)d\Omega^2_{7-p-n}  \, ,\nonumber\\
\phi &=& \phi(u)\, .\nonumber
\eeqa
Here $d\Omega_n^2$ is the metric on a unit $n$ dimensional sphere wrapped by the flavour brane. Similarly  the perpendicular space has been written in adapted polar  coordinates $(\psi, \varphi^i)$.
We will be considering perturbations of the following form
\beqa
Z^{I}(\xi) &\to &  \delta_{I \psi} \left( z^{(0)I}(u) +\epsilon  z^{(1)I}(u) \right) =   \delta_{I \psi}\left(
\psi(u) + \epsilon e^{-i(\omega x^0 - q x^1)}\Psi(u) \rule{0mm}{4mm} \right)\, ,
\nonumber\\
A_a(\xi) &\to & \delta_{a0} A_0(u) + \epsilon  e^{-i(\omega x^0 - q x^1)}\dA_a(u)\, ,
\eeqa
Notice that with this ansatz, we have $\phi\to \phi^{(0)}(u)$ and $\phi^{(1)} = \phi^{(2)} = 0$ in 
(\ref{expfield}).
The equations of motion for the background profiles $\psi(u), A_a(u)$ are obtained from  
${\cal L}_0$ and give
\beqa
\psi&~ \to ~& 2\partial_u \left( e^{-\phi^{(0)}}\sqrt{\gamma} \psi' \gamma^{uu} G_{\psi\psi} \right) - e^{-\phi^{(0)}}\sqrt{\gamma} \left( \psi'^2 \gamma^{uu} G_{\psi\psi,\psi} + n \gamma^{\theta\theta} G_{\theta\theta,\psi} \right) = 0\, ,
\label{geneqpsi}
\\
A_0&~ \to ~& \partial_u \left(e^{-\phi^{(0)}}\sqrt{{\gamma}} \gamma^{0u} \right) = 0\, .
\label{geneqA0}
\eeqa
From ${\cal L}_2$ we obtain the equations of motion for the perturbations. The transverse
fluctuations $\dA_{\perp}(u)=\dA_{2,3}(u) $ satisfy a decoupled equation
\be
EoM[\dA_\perp] \to
\dA_\perp'' + \partial_u \log\left[ e^{-\phi^{(0)}}\sqrt{\gamma} \gamma ^{22} \gamma ^{uu} \right] \dA_\perp' - \frac{\omega^2 \gamma ^{00} + q^2 \gamma ^{11}}{\gamma ^{uu}}\dA_\perp = 0\, .
\label{eomperp}
\ee
On the other hand the longitudinal fluctuations couple to the profile perturbation
\beqa
EoM[\dA_0]~ &\to &~ \dA_0'' + \partial_u \log\left[ e^{-\phi^{(0)}}\sqrt{\gamma} \gamma^{00} \gamma^{uu} \right] \dA_0' - q \frac{\gamma ^{11}}{\gamma ^{uu}} (q \dA_0 + \omega \dA_1)  +q^2 \frac{\gamma^{11}\gamma^{0u}}{\gamma^{00}\gamma^{uu}} \psi' G_{\psi\psi} \Psi 
\label{eomlong1}\nonumber\\
&&~~~~~~~~~~~~~~~~~~~~~~~~~~~~~~~~~~~~~~~
- \frac{ \gamma^{0u}}{\gamma^{00} \gamma^{uu} } \partial_u \left( \Xi \Psi + \Delta \Psi'     \right)  =0 \, ,\\
EoM[\dA_1]~ &\to &~ \dA_1'' + \partial_u \log\left[ e^{-\phi^{(0)}}\sqrt{\gamma} \gamma ^{11} \gamma ^{uu} \right] \dA_1' - \omega\frac{\gamma ^{00}}{\gamma ^{uu}} (q \dA_0 + \omega \dA_1) + q\omega  \frac{\gamma^{0u}}{\gamma^{uu}}  \psi' G_{\psi\psi} \Psi =0 \, ,\nonumber\\
&& \label{eomlong2}\\
EoM[\Psi] ~&\to & ~\Psi'' + \partial_u \log\left[ e^{-\phi^{(0)}}\sqrt{\gamma} \gamma^{uu} G_{\psi\psi} (1-\psi'\Delta) \right] \Psi' + H(u) \Psi  - \frac{\psi' \gamma^{0u} (\Delta'-\Xi) }{\Delta (1-\psi'\Delta)}\dA_0' \nonumber\\
& & ~~~~~~~~~~~~~~~~~~~~~~~~~ + q \frac{\psi' \gamma^{11}\gamma^{0u}}{\gamma^{uu} (1-\psi'\Delta)} (q \dA_0 + \omega \dA_1)  -  \frac{\psi' \gamma^{0u}}{1-\psi'\Delta}  \dA_0''  =0\, ,
\label{eomscal}
\eeqa
and there is  one constraint from the gauge choice, $\dA_u=0$
\be
EoM[\dA_u] ~\to~   i\left( - \omega \gamma ^{00} \dA_0' + q \gamma ^{11} \dA_1' \right) + i \omega \frac{\gamma^{0u}}{\gamma^{uu}}    \left( \Xi \Psi + \Delta \Psi'    \right) =0 \, ,
\ee
where
\be
\Xi  = \med \left( \gamma^{uu}\psi'^2 G_{\psi\psi,\psi} - n \gamma^{\theta\theta} G_{\theta\theta,\psi}\right) 
~~~;~~~
\Delta = \gamma^{uu}\psi' G_{\psi\psi} \, ,
\ee
and
$H(u)$ is a rather lengthy expression we give here for completeness
\beqa 
H(u)&=&\frac{\partial_u \left(e^{-\phi^{(0)}} \sqrt{\gamma} \gamma^{uu} \psi' \left( G_{\psi\psi,\psi} + \frac{n}{2} \gamma^{\theta\theta} G_{\theta\theta,\psi}ÊG_{\psi\psi} -\med \gamma^{uu}\psi'^2 G_{\psi\psi} G_{\psi\psi,\psi}  \right)  \right)}{e^{-\phi^{(0)}} \sqrt{\gamma} \gamma^{uu} G_{\psi\psi} (1-\psi'^2 \gamma^{uu} G_{\psi\psi})}
\nonumber\\
&&-\frac{ \left( \omega^2 (\gamma^{00} -  \psi'^2 G_{\psi\psi}(\gamma^{00}\gamma^{uu}+ (\gamma^{0u})^2)) + q^2 \gamma^{11} (1- \psi'^2 \gamma^{uu}  G_{\psi\psi} )   \right)}{\gamma^{uu}  (1-\psi'^2 \gamma^{uu} G_{\psi\psi})}\nonumber\\
&&-\frac{ \left(  \frac{n(n-2)}{2}  \left( \gamma^{\theta\theta} G_{\theta\theta,\psi} \right)^2 + n \gamma^{uu} \gamma^{\theta\theta} \psi'^2G_{\psi\psi,\psi}ÊG_{\theta\theta,\psi}  + n\gamma^{\theta\theta}G_{\theta\theta,\psi\psi} \right)  }{2\gamma^{uu} G_{\psi\psi} (1-\psi'^2 \gamma^{uu} G_{\psi\psi})}\nonumber\\
&& -\frac{\left( \psi'^2 G_{\psi\psi,\psi\psi} -\med \gamma^{uu}\left( G_{\psi\psi,\psi} \right)^2  \psi'^4 \right)  }{2G_{\psi\psi} (1-\psi'^2 \gamma^{uu} G_{\psi\psi})}\, . \label{Hfactor}
\eeqa
We work with the gauge invariant field combination $\dZ\equiv -i \, e^{i(\omega x_0-q x_1)} F^{(1)}_{10}=q\dA_0+\omega \dA_1$. From the equations for these gauge perturbations we see that taking   combinations
\beqa
&& q\, EoM[\dA_0] + \omega\, EoM[\dA_1] + i  q\omega \frac{\partial_u \log\left( \gamma^{11}/\gamma^{00} \right)}{\gamma^{00}\omega^2 + \gamma^{11} q^2} EoM[\dA_u]\, ,
 \\
&&-\frac{ q (1-\psi'\Delta)}{\psi' \gamma^{0u}}\, EoM[\Psi] +  \omega  EoM[\dA_1] + \frac{iq\omega}{\gamma^{00}\omega^2 + \gamma^{11} q^2} \left[ \log'\left( e^{-\phi^{(0)}}\sqrt{\gamma} \gamma^{11} \gamma^{uu} \right) - \frac{\Delta'-\Xi}{\Delta} \right] EoM[\dA_u]\, ,
\nonumber\\
\eeqa
allows us to write the equations of motion in terms of $\dZ$ and $\Psi$ as follows
\beqa
\dZ'' + {\textswab A}_1 \dZ' + {\textswab B}_1 \dZ + {\textswab C}_1 \Psi'' + {\textswab D}_1 \Psi' + {\textswab E}_1 \Psi & = & 0  \label{gauge1} \\
\dZ'' + {\textswab A}_2 \dZ' + {\textswab B}_2 \dZ + {\textswab C}_2 \Psi'' + {\textswab D}_2 \Psi' + {\textswab E}_2 \Psi & = & 0   \label{gauge2}
\eeqa
with
\beqa
{\textswab A}_1  & = & \log' \left[ \frac{e^{-\phi^{(0)}}\sqrt{\gamma} \gamma^{11} \gamma^{uu}}{\omega^2+ q^2 \frac{\gamma^{11}}{\gamma^{00}}} \right] \nonumber\\
{\textswab A}_2  & = &  \log' \left[ e^{-\phi^{(0)}}\sqrt{\gamma} \gamma^{11} \gamma^{uu} \right] \frac{\omega^2 \gamma^{00}}{\omega^2 \gamma^{00}+ q^2 \gamma^{11}} +\frac{\Delta'-\Xi}{\Delta}  \frac{q^2 \gamma^{11}}{\omega^2 \gamma^{00}+ q^2 \gamma^{11}}   \nonumber \\
{\textswab B}_1  & = & {\textswab B}_2 =  - \frac{\omega^2 \gamma^{00} + q^2 \gamma^{11}}{\gamma^{uu}}\nonumber \\
{\textswab C}_1  & = & -  \frac{q\, \gamma^{0u}}{\gamma^{00}\gamma^{uu}} \Delta  \nonumber \\
{\textswab C}_2  & = & - \frac{q(1-\psi' \Delta)}{\psi' \gamma^{0u}} \nonumber \nonumber \\
{\textswab D}_1  & = & -  \frac{q\, \gamma^{0u}}{\gamma^{00}\gamma^{uu}} \left( \Xi + \Delta' \right)  - \frac{q\omega^2}{\omega^2 \gamma^{00} + q^2 \gamma^{11}}  \frac{\gamma^{0u}}{\gamma^{uu}}  \Delta \,  \log' \left( \frac{\gamma^{11}}{\gamma^{00}} \right) \nonumber \\
{\textswab D}_2  & = & - \frac{q(1-\psi' \Delta)}{\psi' \gamma^{0u}} \log'\left[ e^{-\phi^{(0)}}\sqrt{\gamma} \gamma^{uu} G_{\psi\psi} (1-\psi'\Delta) \right]  \nonumber \\
&& - \frac{\omega^2 \gamma^{00}}{\omega^2 \gamma^{00}+ q^2 \gamma^{11}}  \frac{q\, \gamma^{0u}}{\gamma^{00} \gamma^{uu}}  \left( \log'\left( e^{-\phi^{(0)}}\sqrt{\gamma} \gamma^{11} \gamma^{uu}  \right) - \frac{\Delta'-\Xi}{\Delta} \right) \Delta  \nonumber \\
{\textswab E}_1  & = & -  \frac{q\, \gamma^{0u}}{\gamma^{00}\gamma^{uu}} \left( \Xi '  -\Delta \frac{\omega^2 \gamma^{00} + q^2 \gamma^{11}}{\gamma^{uu}} \right)  - \frac{q\omega^2}{\omega^2 \gamma^{00} + q^2 \gamma^{11}}  \frac{\gamma^{0u}}{\gamma^{uu}}  \Xi \,  \log' \left( \frac{\gamma^{11}}{\gamma^{00}} \right) \nonumber \\
{\textswab E}_2  & = & \frac{q\, \gamma^{0u}}{\gamma^{00}\gamma^{uu}} \left[ \omega^2 \frac{\gamma^{00}}{\gamma^{uu}} \Delta  - \frac{\omega^2 \gamma^{00}}{\omega^2 \gamma^{00}+ q^2 \gamma^{11}} \, \Xi  \left( \log'\left( e^{-\phi^{(0)}}\sqrt{\gamma} \gamma^{11} \gamma^{uu}  \right) - \frac{\Delta'-\Xi}{\Delta} \right) \right] - \frac{q(1-\psi' \Delta)}{\psi' \gamma^{0u}} H(u) \nonumber
\eeqa
In the particular case of the D3/D7 intersection, we can express the factors governing the equations of motion as
\beqa
&&\gamma_{00}=-f(u)\frac{(\pi T L)^2}{u} ~~;~~\gamma_{11}=\gamma_{22}=\gamma_{33}=\frac{(\pi T L)^2}{u}\, , \nonumber\\
&&\gamma_{uu}= \frac{L^2(1-\psi^2+4u^2f(u)\psi'^2)}{4(1-\psi^2)u^2f(u)}~~;~~\gamma_{0u}= -\gamma_{u0}=-2\pi\alpha' A_0'(u)\, , \label{d3d7}\\
&&\gamma_{\theta\theta}=G_{\theta\theta}= L^2 (1-\psi^2) ~~;~~G_{\psi\psi}=\frac{L^2}{1-\psi^2} ~~;~~n=3~~;~~\phi^{(0)}=0\, .\nonumber
\eeqa
with inverse components
\beqa
&&\gamma^{00}=\frac{\gamma_{uu}}{\gamma_{00}\gamma_{uu}+(\gamma_{0u})^2} ~~;~~\gamma^{11}=\gamma^{22}=\gamma^{33}=\frac{1}{\gamma_{11}}\, ~~;~~\gamma^{\theta\theta}=\frac{1}{\gamma_{\theta\theta}}, \label{d3d7inv}\\
&&\gamma^{uu}= \frac{\gamma_{00}}{\gamma_{00}\gamma_{uu}+(\gamma_{0u})^2}~~;~~\gamma^{0u}= -\gamma^{u0}=\frac{-\gamma_{0u}}{\gamma_{00}\gamma_{uu}+(\gamma_{0u})^2} \nonumber \, .
\eeqa

\section{Boundary conditions for  the coupled system \label{apbound}}

In this paper we deal with a coupled system of second order differential equations with a regular singular point at $p=0$. 
That is 
\beqa
Z''(p) + \frac{A(p)}{p} Z'(p)+ \frac{B(p)}{p^2} Z(p)+ \frac{C(p)}{p} \Psi'(p)+ \frac{D(p)}{p^2} \Psi(p) & = & 0\, , 
\label{canonical2}\\
\Psi''(p) + \frac{\tilde A(p)}{p} \Psi'(p)+ \frac{\tilde B(p)}{p^2} \Psi(p) +  \frac{\tilde C(p)}{p} Z'(p)+ \frac{\tilde D(p)}{p^2} Z(p) & = & 0 \, ,
\label{canonical1}
\eeqa
where the singularity  at $p=0$ is such that all the functions $\{A(p),\, B(p),\dots \tilde C(p),\, \tilde D(p)\}$ are regular and can be Taylor-expanded in powers of $p$. This expansion will be indicated in general as
\be
M(p)=\sum_{i=0}^\infty m_i p^i\, ,
\ee
i.e., using small letters.
Performing a Frobenius expansion for $Z$ and $\Psi$ with indices $\lambda$ and $\eta$
\be
 Z=p^\lambda \sum\limits_{i=0}^\infty z_i p^i ~~;~~  \Psi=p^\eta \sum\limits_{i=0}^\infty \psi_i p^i  \, ,
\ee
 a system of recursion relations for the coefficients $z_i$ and $\psi_i$ can be calculated
\beqa
\frac{1}{p^2} \sum\limits_{k=0}^\infty \sum\limits_{i=0}^k \left[  p^{\lambda} \left( (\lambda+i)(\lambda+i-1)\delta_{ik} + (\lambda+i) a_{k-i} + b_{k-i} \right) z_i + p^\eta \left( (\eta+i) c_{k-i} + d_{k-i} \right) \psi_i  \right] p^k & = & 0\, , \nonumber \\
\frac{1}{p^2} \sum\limits_{k=0}^\infty \sum\limits_{i=0}^k \left[  p^{\eta} \left( (\eta+i)(\eta+i-1)\delta_{ik} + (\eta+i) \tilde a_{k-i} + \tilde b_{k-i} \right) \psi_i + p^\lambda \left( (\lambda+i) \tilde c_{k-i} + \tilde d_{k-i} \right) z_i  \right] p^k & = & 0\, . \nonumber
\eeqa

We can now focus on the D3/D7  background.  Close to the horizon we take $p=1-u$. For equations (\ref{gauge2}) and (\ref{gauge1}) we obtain the following coefficients
\be
a_0=\tilde a_0 = 1 ~~;~~~~
b_0 = \tilde b_0 = \frac{\omega^2}{16\pi^2 T^2} ~~;~~~~
c_0 =  \tilde c_0 = 0 ~~;~~~~
d_0 =  \tilde d_0 = 0\, ,
\ee

and find  two solutions of the indicial equation 
\be
\lambda_{\pm}=\pm i \frac{\omega}{4\pi T}=\eta_{\pm}\, .
\ee
There is one free parameter, say $z_0$ (or $\psi_0$). Now for an analysis close to the boundary let $p=u$ in which case, we get
\be
\begin{array}{ccc}
a_0 & = & 0\, ,\\
\tilde a_0 & = & -1 \rule{0mm}{5mm} \\
\end{array}~~~;~~~
\begin{array}{ccc}
b_0  & = & 0  \\
\tilde b_0 & = &\displaystyle  \frac{3}{4} \, \\
\end{array}~~~;~~~
\begin{array}{rcl}
c_0 = d_0& = & 0\, \\
 \tilde c_0  =  \tilde d_0 & = & 0  \rule{0mm}{6mm} \, ,
\end{array}
\ee
giving the following exponents
\be
\lambda_{1,2}=\{0,\,1\}\,\, \, \, \, ; \hspace{1cm} \eta_{1,2}=\bigg\{\med,\,\frac{3}{2}\bigg\}\, .
\ee

 With these coefficients we find near the boundary
\beqa
Z(u) & = & {\mathcal A}\, Z^{(A)}(u) + u\, {\mathcal B}\,  Z^{(B)}(u)\, , \\
\Psi(u) & = & u^{1/2}\, {\mathcal M}\,  \Psi^{(M)}(u) + u^{3/2}\, {\mathcal C}\,  \Psi^{(C)}(u)\, ,
\eeqa
where
\beqa
Z^{(A)}(u) & = & 1+\sum\limits_{i=1}^\infty z^{(a)}_i u^i - \frac{\omega^2-q^2}{4\pi^2 T^2} u \log\, u \, Z^{(B)}(u)~~~;~~~
Z^{(B)}(u)  =  1+ \sum\limits_{i=1}^\infty z^{(b)}_i u^i\, , \\
\Psi^{(M)}(u) & = & 1+\sum\limits_{i=1}^\infty \psi^{(m)}_i u^i -   \frac{\omega^2-q^2}{4\pi^2 T^2} u \log\, u \, \Psi^{(C)}(u)~~~;~~~
\Psi^{(C)}(u) =  1+ \sum\limits_{i=1}^\infty \psi^{(c)}_i u^i \, .
\eeqa

\section{Asymptotic expressions for lightlike momenta \label{asymptote}}

In order to trust the numerical results it is important to compare both to previous results, or to analytical expressions wherever possible. In the current setup it is possible to calculate such analytic results only in certain asymptotic limits. Setting the momenta on the light cone, $\wn=\qn$,  the longitudinal polarization vanishes and the equation for the transverse component of the gauge field (\ref{eomperp}) acquires the following form upon introducing the particular values given in (\ref{d3d7}) and (\ref{d3d7inv})
\begin{equation}
{\cal A}_\perp '' +
 \partial_u \log\left(f\,\tpsi\sqrt{\frac{\tpsi^6+\Dis^2u^3}{\tpsi^2+4u^2f\psi'^2}}\right)\, {\cal A}_\perp ' + 
\frac{\wn^2}{u\,f^2} \frac{\tpsi^6(1-f)+\Dis^2 u^3}{\tpsi^2(\tpsi^6+\Dis^2u^3)} \, {\cal A}_\perp =0 \, . \label{transv}
\end{equation}

In contrast to the massless, $D=0$ case \cite{hep-th/0607237}, we have found no analytic
solution when baryon number is turned on. We can however perform analytic expansions in the small and large $\wn$ regimes and extract analytic information perturbatively.

\subsection{Low frequency limit}

For the small $\wn$ limit we will use a perturbative approach to find the analytic behaviour. We begin extracting the regular singularity at $u=1$ by substituting
\begin{equation}
{\cal A}_\perp(u)=f^{-i\wn /2} \left(Y_0(u)+ \wn Y_1(u) + {\cal O}(\wn^2)\right)\, ,
\end{equation}
with $Y_n(u)$ regular at $u=1$. We can then expand the equation of motion as a series in $\wn$ and solve it order by order. From \eqref{transv} we are led to the following equation for $Y_0(u)$, 
\begin{eqnarray}
Y_0'=\frac{C_1}{f\,\tpsi}\sqrt{\frac{\tpsi^2+4u^2f\psi'^2}{\tpsi^6+\Dis^2u^3}}\ \, .
\end{eqnarray}
Clearly to avoid a singular solution at $u=1$ we must set $C_1=0$ and hence $Y_0$ is a constant. For $Y_1$ we perform the same procedure, finding the general solution
\be
Y_1'=\frac{C_2}{f\,\tpsi}\sqrt{\frac{\tpsi^2+4u^2f\psi'^2}{\tpsi^6+\Dis^2u^3}}+\frac{i}{2}\partial_u\left( \log (f) Y_0 \right)\, ,
\ee
which is regularized at the horizon setting
\be
C_2 = i\,Y_0\sqrt{(1-\psi_0^2)^3+D^2}\, ,
\ee
where $\psi_0=\psi(u=1)$.

We can calculate the spectral function by looking at the $u\rightarrow 0$ end of the solution and find that
\begin{eqnarray} \label{low-Spectral}
\chi^\mu_\mu \sim \frac{N_c N_f T^2}{2} \wn\sqrt{(1-\psi_0^2)^3+\Dis^2}+{\cal O}(\wn^2)\, ,
\end{eqnarray}
which leads to precisely the same result for the conductivity as that obtained in the vanishing electric field limit in \cite{arXiv:0705.3870}. This is a very elegant result showing that the microscopic and macroscopic calculations agree with each other.

\subsection{High frequency limit}

Now we move to the $\wn \gg 1$ limit. We use the Langer-Olver method (a version of the WKB approximation) to construct the
asymptotic solution and will consider only the massless case.
Following \cite{hep-th/0607237,arXiv:0802.1460}, we perform the following
transformation
\begin{equation}
{\cal A}_\perp(u)=\sqrt{\frac{1}{f}\sqrt{\frac{1}{1+\Dis^2u^3}}} \:y(u) \, \, ; \quad u=-x \, ,
\end{equation}
for the equation \eqref{transv}, in order to rewrite it in a
Schr\"odinger form
\begin{equation}
y''(x)=[{\wn }^2 H(x)+G(x)]y(x) \label{largeomega}\, ,
\end{equation}
where
\begin{equation} H(x)=\frac{\wn^2}{x\,f(x)^2} \frac{1-f(x)-\Dis^2 x^3}{1-\Dis^2x^3}\, ,
\end{equation}
$x=-u\in[-1,0]$, and $G(x)$ is another function which will not be needed in the following. For large
$\wn$ the dominant term on the right-hand side of
\eqref{largeomega} has a simple zero at $x=0$ and thus, according
to \cite{olver}, the asymptotic solution can be expressed in terms
of Airy functions. With this goal in mind, we change variables from
$x$ to $\zeta$ as
\begin{equation}
\zeta\left(\frac{d\zeta}{dx}\right)^2= H(x) \quad \Rightarrow
\quad \zeta=\left[-\frac{3}{2}\int_0^x\sqrt{H(t)}dt\right]^{2/3} \, , \end{equation}
and rescale $y(x)$ to $W(\zeta)$ as
\begin{eqnarray}
y&=&\left(\frac{d\zeta}{dx}\right)^{-1/2}W \, .
\end{eqnarray}
After which we are left with a new differential equation for
$W(\zeta)$
\begin{equation}
\frac{d^2W}{d\zeta^2}=[\wn ^2\zeta+\eta(\zeta)]W \, ,
\label{Airy}
\end{equation}
where $\eta(\zeta)$ is another function whose form is not needed in the analysis\footnote{The exact
expression for this function in terms of $G(x)$ and $H(x)$ is in
\cite{arXiv:0802.1460}}.
For large $\wn $ \eqref{Airy} reduces to Airy's equation and
the solution to leading order is
\begin{equation}
W(\zeta)=A_0\mbox{Ai}(\wn ^{2/3}\zeta)+B_0\mbox{Bi}
(\wn ^{2/3}\zeta)+\ldots\, .
\end{equation}
Dictated by the incoming-wave boundary conditions at the horizon
we set $B_0$ to zero. Thus the approximate solution for
${\cal A}_\perp(u)$ for large $\wn $ is
\begin{equation}
{\cal A}_\perp(u)=A_0
\left[\frac{\zeta(-u)}{u(1+D^2u)}\right]^{1/4} \mbox{Ai}(\omega^{2/3}\zeta(-u))+\ldots \, ,
\end{equation}
leading to the following expression for the transverse correlator
\begin{equation} \label{TransverseScalar}
\Pi^\perp =
-\frac{N_c N_fT^2}{8}\left(-\frac{\Dis^2}{5}+\frac{(-1)^{2/3}3^{1/3}\Gamma(2/3)}{\Gamma(1/3)}\wn ^{2/3}\right)\, .
\end{equation}
Note that the transverse correlator depends on $D$, but only through
the real part. The trace of the spectral function in the
high-frequency limit for lightlike momenta is
\footnote{In the massless case the differential equation for the
scalar mode is decoupled from the longitudinal one, and
its contribution to the spectral function can be calculated
independently. In the lightlike case, even if the initial
differential equation for the scalar is different from that for
${\cal A}_\perp$, we can prove that the contributions to leading order in
$\omega$ in the transverse scalar and consequently to the spectral
function are \eqref{TransverseScalar} and \eqref{Spectral}.}
\begin{eqnarray} \label{Spectral}
\chi^\mu_\mu \sim \frac{N_c N_fT^2}{2}\,
\frac{3^{5/6}\Gamma(2/3)}{2\,\Gamma(1/3)} \wn^{2/3} +{\cal O}(\wn)\, ,
\end{eqnarray}
and is independent of the presence of baryon density, at least to
leading order in $\wn$, giving the same result as in
\cite{hep-th/0607237}. The appearance of the baryon density in the real
part of the transverse scalar indicates possible deviation of our
result from that of \cite{hep-th/0607237} at the first subleading order in
$\wn$. The numerical solution matches this claim.

\section{The Wess-Zumino  term}\label{apWZ}

In order to illustrate that there is no contribution from the WZ term to the equations of motion, we start by writing the Ramond-Ramond four-form $C_4$ in the following notation:
\begin{equation} \label{4form}
C_4=-\frac{(\pi T L)^4}{u^2} dt\wedge dx_1 \wedge dx_2 \wedge dx_3\, .
\end{equation}
The contribution of this four-form to the five-form field strength is:
\begin{equation}
dC_4=\frac{(\pi T L)^4}{u^3} dt\wedge dx_1 \wedge dx_2 \wedge dx_3 \wedge du\, ,
\end{equation}
whereas the Hodge-dual contribution is:
\begin{equation}
\star dC_4=4L^5(1-\psi^2)\psi \sin\theta\cos\theta d\psi\wedge d\phi \wedge d\theta \wedge d\phi_2\wedge d\phi_3 \, .
\end{equation}
From this expression we can deduce the four-form dual to the one in \eqref{4form}:
\begin{eqnarray} \label{4form-hodge}
\tilde{C}_4&=&\frac{2L^5}{3}\Bigg[(1-\psi^2)\psi\left(\sin^2\theta d\psi\wedge d\phi-\sin 2\theta \, \phi \, d\psi\wedge d\theta\right)  \nonumber \\ 
&+&\frac{\psi^2}{2}\left(1-\frac{\psi^2}{2}\right)\sin 2\theta d\phi\wedge d\theta\Bigg]d\phi_2\wedge d\phi_3\, . 
\end{eqnarray}
Next, we use \eqref{4form-hodge} to calculate its pullback onto the D7-brane, so we have:
\begin{eqnarray}
P\left[\tilde{C}_4\right]&=&\frac{2L^5}{3}\left(1-\psi^2\right)\psi\sin^2\theta\frac{\partial\psi}{\partial x^\alpha}\frac{\partial\phi}{\partial x^\beta}dx^\alpha\wedge dx^\beta \wedge d\phi_2 \wedge d\phi_3\nonumber\\
&+&\frac{2L^5}{3}\left(1-\psi^2\right)\psi\phi\sin 2\theta\frac{\partial\psi}{\partial x^\alpha}dx^\alpha\wedge d\theta \wedge d\phi_2 \wedge d\phi_3\nonumber\\
&+&\frac{L^5}{3}\psi^2\left(1-\frac{\psi^2}{2}\right)\sin 2\theta\frac{\partial\phi}{\partial x^\alpha}dx^\alpha\wedge d\theta \wedge d\phi_2 \wedge d\phi_3\, .
\end{eqnarray}
We are interested in WZ terms with a product of two fields which are fluctuations on the D7-brane. Each term in the above four-form will contain at least one factor of a fluctuating field. The pull-back must be contracted with two factors of the brane world-volume gauge field, and therefore one of these two factors must be the background component $F_{0u}$ corresponding to the presence of the finite baryon number density. The other component will therefore be a fluctuation of a gauge field. The eight-form product which will contribute at second order will have the following three terms:
\begin{eqnarray}
F\wedge F\wedge P\left[\tilde{C}_4\right]&=& F_{0u}\wedge F_{12}\wedge \left(L^5\psi^2\left(1-\frac{\psi^2}{2}\right)\sin 2\theta\frac{\partial\phi}{\partial x^3}\right)dx^3\wedge d\theta \wedge d\phi_2 \wedge d\phi_3\nonumber\\
&+& F_{0u}\wedge F_{23} \left(L^5\psi^2\left(1-\frac{\psi^2}{2}\right)\sin 2\theta\frac{\partial\phi}{\partial x^1}\right)dx^1\wedge d\theta \wedge d\phi_2 \wedge d\phi_3\nonumber\\
&+& F_{0u}\wedge F_{31} \left(L^5\psi^2\left(1-\frac{\psi^2}{2}\right)\sin 2\theta\frac{\partial\phi}{\partial x^2}\right)dx^2\wedge d\theta \wedge d\phi_2 \wedge d\phi_3\, .\nonumber\\ 
\end{eqnarray}
This can be written as:
\begin{eqnarray}
-A_{0}'L^5 \psi^2\left(1-\frac{\psi^2}{2}\right)\sin 2 \theta \left[(\partial_1A_2-\partial_2 A_1)\frac{\partial\phi}{\partial x_3}  \: \text{+ cyclic in 1,2,3}\right]\text{Vol}_{D7} \, .
\end{eqnarray}
Then it is easy  to produce the WZ contribution to the equation of motion for $\phi$:
\begin{equation}
-A_{0}'L^5 \psi\left(1-\frac{\psi^2}{2}\right)\sin 2\theta \left[\partial_3\left(\partial_1A_2-\partial_2A_1\right)+\partial_1\left(\partial_2A_3-\partial_3A_2\right)+\partial_2\left(\partial_3A_1-\partial_1A_3\right)\right]\, .
\end{equation}
From this equation it is obvious that the terms coming from the WZ part of the action vanish due to the antisymmetry of the gauge field indices. 

For the field $A_1$ the WZ contribution to the equation of motion is given by:
\begin{equation}
\frac{\partial}{\partial x_2}\left[A_{0}' L^5\psi\left(1-\frac{\psi^2}{2}\right)\sin 2\theta\frac{\partial\phi}{\partial x_3}\right]-\frac{\partial}{\partial x_3}\left[A_{0}' L^5\psi\left(1-\frac{\psi^2}{2}\right)\sin 2\theta\frac{\partial\phi}{\partial x_2}\right]\, ,
\end{equation}
which once again gives no contribution from the WZ term. For the fields $A_2$ and $A_3$ the above argument goes through identically.

\section{Comparison with the conductivity found by Karch and O'Bannon}\label{ap.KarchOB}

In \cite{arXiv:0705.3870} the following result for the conductivity was obtained:

\begin{equation}
\tilde\sigma=\sqrt{\frac{N_f^2N_c^2T^2}{16\pi^2}\cos^6\theta+\frac{\tilde \Dis^2}{\left(\frac{\pi}{2}\right)^2\lambda
T^4}}\, .
\end{equation}
The difference in normalisation between our conductivity, $\sigma$, and theirs, $\tilde\sigma$, is:
\begin{equation}
\tilde \sigma=\frac{4}{e^2}\sigma\frac{N_fN_c T}{4\pi}\, .
\end{equation}
The factor $\frac{N_fN_c T}{4\pi}$ comes from the normalisation of $\chi$ and the factor
$\frac{4}{e^2}$ is a difference in definition between our
conductivities.
Therefore:
 \begin{equation}
\frac{4}{e^2}\sigma=\sqrt{(1-\psi_0^2)^3+\frac{16\pi^2}{N_fN_c}\frac{\tilde\Dis^2}{\left(\frac{\pi}{2}\right)^2\lambda
T^4}}\, .
\end{equation}

Inserting their definition of $\lambda$ and remembering to put back in
the AdS radius gives:
 \begin{equation}
\sigma =\frac{e^2}{4}\sqrt{(1-\psi_0^2)^3+\frac{\tilde\Dis^2 64
\alpha'^2}{T^6 N_f^2 N_c^2 L^4}}\, .
\end{equation}
The difference between our definitions of $D$ is calculated by looking
at the UV behaviour of $A_{0}$. We must also note that the difference in
our definitions of the radial coordinate are given by:
\begin{equation}
\frac{z}{z_H}=\frac{\sqrt{u}}{\sqrt{\sqrt{1-u^2}+1}}\, ,
\label{changezu}
\end{equation}
where $z_H=\frac{\sqrt{2}}{\pi T}$.

This gives a difference in $D$s as:
\begin{equation}
\tilde\Dis=\frac{N_fN_c T^3 L^2}{8\alpha'}\Dis\, ,
\end{equation}
where again the tilded quantity is their definition.
Using this in the definition of the conductivity gives us:
\begin{equation}
\sigma=\frac{e^2}{4}\sqrt{(1-\psi_0^2)^3+\Dis^2}\, .
\end{equation}

This result matches perfectly with our numerical calculation given in section \ref{conductivitysec}.





\begin{thebibliography}{99}

\bibitem{hep-th/9711200}
 J.~M.~Maldacena,
 ``The large N limit of superconformal field theories and supergravity,''
 Adv.\ Theor.\ Math.\ Phys.\  {\bf 2}, 231 (1998)
 [Int.\ J.\ Theor.\ Phys.\  {\bf 38}, 1113 (1999)]
 [arXiv:hep-th/9711200].

\bibitem{hep-th/9905111}
 O.~Aharony, S.~S.~Gubser, J.~M.~Maldacena, H.~Ooguri and Y.~Oz,
 ``Large N field theories, string theory and gravity,''
 Phys.\ Rept.\  {\bf 323}, 183 (2000)
 [arXiv:hep-th/9905111].

\bibitem{arXiv:0804.2423}
 R.~C.~Myers and S.~E.~Vazquez,
 ``Quark Soup al dente: Applied Superstring Theory,''
 arXiv:0804.2423 [hep-th].

\bibitem{hep-th/9803131}
  E.~Witten,
 ``Anti-de Sitter space, thermal phase transition, and confinement in  gauge
  theories,''
  Adv.\ Theor.\ Math.\ Phys.\  {\bf 2}, 505 (1998)
  [arXiv:hep-th/9803131].

\bibitem{hep-th/0602059}
 P.~Kovtun and A.~Starinets,
 ``Thermal spectral functions of strongly coupled N = 4 supersymmetric
 Yang-Mills theory,''
 Phys.\ Rev.\ Lett.\  {\bf 96}, 131601 (2006)
 [arXiv:hep-th/0602059].

\bibitem{arXiv:0706.0162}
 R.~C.~Myers, A.~O.~Starinets and R.~M.~Thomson,
 ``Holographic spectral functions and diffusion constants for fundamental
 matter,''
 JHEP {\bf 0711}, 091 (2007)
 [arXiv:0706.0162 [hep-th]].

\bibitem{arXiv:0709.2168}
 D.~Mateos and L.~Patino,
 ``Bright branes for strongly coupled plasmas,''
 JHEP {\bf 0711}, 025 (2007)
 [arXiv:0709.2168 [hep-th]].

\bibitem{arXiv:0710.0334}
 J.~Erdmenger, M.~Kaminski and F.~Rust,
 ``Holographic vector mesons from spectral functions at finite baryon or
 isospin density,''
 Phys.\ Rev.\  D {\bf 77}, 046005 (2008)
 [arXiv:0710.0334 [hep-th]].
 
 \bibitem{arXiv:0709.3948}
  O.~Aharony, K.~Peeters, J.~Sonnenschein and M.~Zamaklar,
  ``Rho meson condensation at finite isospin chemical potential in a
  holographic model for QCD,''
  JHEP {\bf 0802}, 071 (2008)
  [arXiv:0709.3948 [hep-th]].



\bibitem{Amado:2007yr}
  I.~Amado, C.~Hoyos-Badajoz, K.~Landsteiner and S.~Montero,
 ``Residues of Correlators in the Strongly Coupled N=4 Plasma,''
  Phys.\ Rev.\  D {\bf 77}, 065004 (2008)
  [arXiv:0710.4458 [hep-th]].

\bibitem{hep-th/0405231}
 P.~Kovtun, D.~T.~Son and A.~O.~Starinets,
 ``Viscosity in strongly interacting quantum field theories from black hole
 physics,''
 Phys.\ Rev.\ Lett.\  {\bf 94}, 111601 (2005)
 [arXiv:hep-th/0405231].

\bibitem{McLerran:1984ay}
  L.~D.~McLerran and T.~Toimela,
  ``Photon And Dilepton Emission From The Quark - Gluon Plasma: Some General
  Considerations,''
  Phys.\ Rev.\  D {\bf 31}, 545 (1985).

\bibitem{hep-th/0607237}
 S.~Caron-Huot, P.~Kovtun, G.~D.~Moore, A.~Starinets and L.~G.~Yaffe,
 ``Photon and dilepton production in supersymmetric Yang-Mills plasma,''
 JHEP {\bf 0612}, 015 (2006)
 [arXiv:hep-th/0607237].

\bibitem{hep-th/0205236}
 A.~Karch and E.~Katz,
 ``Adding flavor to AdS/CFT,''
 JHEP {\bf 0206}, 043 (2002)
 [arXiv:hep-th/0205236].

\bibitem{hep-th/0306018}
 J.~Babington, J.~Erdmenger, N.~J.~Evans, Z.~Guralnik and I.~Kirsch,
 ``Chiral symmetry breaking and pions in non-supersymmetric gauge /  gravity
 duals,''
 Phys.\ Rev.\  D {\bf 69}, 066007 (2004)
 [arXiv:hep-th/0306018].

\bibitem{hep-th/0611021}
 S.~Nakamura, Y.~Seo, S.~J.~Sin and K.~P.~Yogendran,
 ``A new phase at finite quark density from AdS/CFT,''
 arXiv:hep-th/0611021.

\bibitem{hep-th/0611099}
 S.~Kobayashi, D.~Mateos, S.~Matsuura, R.~C.~Myers and R.~M.~Thomson,
 ``Holographic phase transitions at finite baryon density,''
 JHEP {\bf 0702}, 016 (2007)
 [arXiv:hep-th/0611099].




\bibitem{hep-th/0610247}
 A.~Parnachev and D.~A.~Sahakyan,
 ``Photoemission with chemical potential from QCD gravity dual,''
 Nucl.\ Phys.\  B {\bf 768}, 177 (2007)
 [arXiv:hep-th/0610247].

\bibitem{hep-th/0412141}
  T.~Sakai and S.~Sugimoto,
  ``Low energy hadron physics in holographic QCD,''
  Prog.\ Theor.\ Phys.\  {\bf 113}, 843 (2005)
  [arXiv:hep-th/0412141].

\bibitem{arXiv:0802.1460}
 A.~Nata Atmaja and K.~Schalm,
 ``Photon and Dilepton Production in Soft Wall AdS/QCD,''
 arXiv:0802.1460 [hep-th].

\bibitem{hep-ph/0509068}
 T.~Schafer,
 ``Phases of QCD,''
 arXiv:hep-ph/0509068.

\bibitem{hep-ph/0011333}
 K.~Rajagopal and F.~Wilczek,
 ``The condensed matter physics of QCD,''
 arXiv:hep-ph/0011333.


\bibitem{arXiv:0705.3870}
 A.~Karch and A.~O'Bannon,
 ``Metallic AdS/CFT,''
 JHEP {\bf 0709}, 024 (2007)
 [arXiv:0705.3870 [hep-th]].




\bibitem{arXiv:0708.2818}
 S.~Nakamura, Y.~Seo, S.~J.~Sin and K.~P.~Yogendran,
 ``Baryon-charge Chemical Potential in AdS/CFT,''
 arXiv:0708.2818 [hep-th].

\bibitem{arXiv:0708.3706}
 K.~Ghoroku, M.~Ishihara and A.~Nakamura,
 ``D3/D7-holographic Gauge theory and Chemical potential,''
 Phys.\ Rev.\  D {\bf 76}, 124006 (2007)
 [arXiv:0708.3706 [hep-th]].

\bibitem{arXiv:0709.0570}
 A.~Karch and A.~O'Bannon,
 ``Holographic Thermodynamics at Finite Baryon Density: Some Exact Results,''
 JHEP {\bf 0711}, 074 (2007)
 [arXiv:0709.0570 [hep-th]].

\bibitem{arXiv:0709.1225}
 D.~Mateos, S.~Matsuura, R.~C.~Myers and R.~M.~Thomson,
 ``Holographic phase transitions at finite chemical potential,''
 JHEP {\bf 0711}, 085 (2007)
 [arXiv:0709.1225 [hep-th]].

\bibitem{arXiv:0711.0407}
 S.~Matsuura,
 ``On holographic phase transitions at finite chemical potential,''
 JHEP {\bf 0711}, 098 (2007)
 [arXiv:0711.0407 [hep-th]].




\bibitem{arXiv:0706.2191}
  L.~McLerran and R.~D.~Pisarski,
 ``Phases of Cold, Dense Quarks at Large $N_c$,''
  Nucl.\ Phys.\  A {\bf 796}, 83 (2007)
  [arXiv:0706.2191 [hep-ph]].

\bibitem{arXiv:0803.0279}
 Y.~Hidaka, L.~D.~McLerran and R.~D.~Pisarski,
 ``Baryons and the phase diagram for a large number of colors and flavors,''
 arXiv:0803.0279 [hep-ph].

\bibitem{nucl-th/0511071}
 A.~Andronic, P.~Braun-Munzinger and J.~Stachel,
 ``Hadron production in central nucleus nucleus collisions at chemical
 freeze-out,''
 Nucl.\ Phys.\  A {\bf 772}, 167 (2006)
 [arXiv:nucl-th/0511071].

\bibitem{hep-ph/0511094}
 J.~Cleymans, H.~Oeschler, K.~Redlich and S.~Wheaton,
 ``Comparison of chemical freeze-out criteria in heavy-ion collisions,''
 Phys.\ Rev.\  C {\bf 73}, 034905 (2006)
 [arXiv:hep-ph/0511094].






\bibitem{arXiv:0804.2168}
 R.~C.~Myers and A.~Sinha,
 ``The fast life of holographic mesons,''
 arXiv:0804.2168 [hep-th].
\bibitem{hep-th/0612169}
  C.~Hoyos-Badajoz, K.~Landsteiner and S.~Montero,
  ``Holographic Meson Melting,''
  JHEP {\bf 0704}, 031 (2007)
  [arXiv:hep-th/0612169].









\bibitem{hep-th/0311270}
 M.~Kruczenski, D.~Mateos, R.~C.~Myers and D.~J.~Winters,
 ``Towards a holographic dual of large-N(c) QCD,''
 JHEP {\bf 0405}, 041 (2004)
 [arXiv:hep-th/0311270].

\bibitem{hep-th/0304032}
 M.~Kruczenski, D.~Mateos, R.~C.~Myers and D.~J.~Winters,
 ``Meson spectroscopy in AdS/CFT with flavour,''
 JHEP {\bf 0307}, 049 (2003)
 [arXiv:hep-th/0304032].

\bibitem{hep-th/0602174}
 D.~Arean and A.~V.~Ramallo,
 ``Open string modes at brane intersections,''
 JHEP {\bf 0604}, 037 (2006)
 [arXiv:hep-th/0602174].

\bibitem{hep-th/0605261}
 A.~V.~Ramallo,
 ``Adding open string modes to the gauge / gravity correspondence,''
 Mod.\ Phys.\ Lett.\  A {\bf 21}, 1481 (2006)
 [arXiv:hep-th/0605261].

\bibitem{hep-th/0605017}
 R.~C.~Myers and R.~M.~Thomson,
 ``Holographic mesons in various dimensions,''
 JHEP {\bf 0609}, 066 (2006)
 [arXiv:hep-th/0605017].

\bibitem{hep-th/0701132}
 D.~Mateos, R.~C.~Myers and R.~M.~Thomson,
 ``Thermodynamics of the brane,''
 JHEP {\bf 0705}, 067 (2007)
 [arXiv:hep-th/0701132].

\bibitem{hep-th/0605046}
 D.~Mateos, R.~C.~Myers and R.~M.~Thomson,
 ``Holographic phase transitions with fundamental matter,''
 Phys.\ Rev.\ Lett.\  {\bf 97}, 091601 (2006)
 [arXiv:hep-th/0605046].

\bibitem{arXiv:0711.4467}
 J.~Erdmenger, N.~Evans, I.~Kirsch and E.~Threlfall,
 ``Mesons in Gauge/Gravity Duals - A Review,''
 Eur.\ Phys.\ J.\  A {\bf 35}, 81 (2008)
 [arXiv:0711.4467 [hep-th]].

\bibitem{hep-th/0205051}
 D.~T.~Son and A.~O.~Starinets,
 ``Minkowski-space correlators in AdS/CFT correspondence: Recipe and
 applications,''
 JHEP {\bf 0209}, 042 (2002)
 [arXiv:hep-th/0205051].

\bibitem{arXiv:0803.0759}
 A.~Paredes, K.~Peeters and M.~Zamaklar,
 ``Mesons versus quasi-normal modes: undercooling and overheating,''
 arXiv:0803.0759 [hep-th].


\bibitem{arXiv:0712.0590}
 Q.~J.~Ejaz, T.~Faulkner, H.~Liu, K.~Rajagopal and U.~A.~Wiedemann,
 ``A limiting velocity for quarkonium propagation in a strongly coupled plasma
 via AdS/CFT,''
 JHEP {\bf 0804}, 089 (2008)
 [arXiv:0712.0590 [hep-th]].

\bibitem{Fox:1970cu}
 R.~Fox, C.~G.~Kuper and S.~G.~Lipson,
 ``Faster-than-light group velocities and causality violation,''
 Proc.\ Roy.\ Soc.\ Lond.\  {\bf 316}, 515 (1970).

\bibitem{Fox:1969us}
 R.~Fox, C.~G.~Kuper and S.~G.~Lipson,
 ``Do faster-than-light group velocities imply violation of causality?,''
 Nature {\bf 223}, 597 (1969).



\bibitem{arXiv:0802.0775}
  N.~Evans and E.~Threlfall,
 ``Mesonic quasinormal modes of the Sakai-Sugimoto model at high
  temperature,''
  arXiv:0802.0775 [hep-th].

\bibitem{olver}
  F.~W.~J.~Olver, {\it Asymptotics and special functions},
  A\, K\, Peters, Wellesley, 1997.


\end{thebibliography}
\end{document}